\documentclass{emulateapj}
\usepackage{ifthen}

\newcommand{\ctbd}[1]{}

\newcommand{\lc}{light curve}
\newcommand{\lcs}{light curves}
\newcommand{\Lc}{Light curve}

\newcommand{\oot}{out-of-transit}

\newcommand{\band}[1]{\ensuremath{#1}-band}

\newcommand{\kms}{\ensuremath{\rm km\,s^{-1}}}
\newcommand{\ms}{\ensuremath{\rm m\,s^{-1}}}

\newcommand{\gcmc}{\ensuremath{\rm g\,cm^{-3}}}
\newcommand{\ergscmsq}{\ensuremath{\rm erg\,s^{-1}\,cm^{-2}}}

\newcommand{\teff}{\ensuremath{T_{\rm eff}}}

\newcommand{\vsini}{\ensuremath{v \sin{i}}}
\newcommand{\feh}{\ensuremath{\rm [Fe/H]}}

\newcommand{\vmac}{\ensuremath{v_{\rm mac}}}
\newcommand{\vmic}{\ensuremath{v_{\rm mic}}}

\newcommand{\rsun}{\ensuremath{R_\sun}}
\newcommand{\msun}{\ensuremath{M_\sun}}
\newcommand{\lsun}{\ensuremath{L_\sun}}

\newcommand{\rstar}{\ensuremath{R_\star}}
\newcommand{\mstar}{\ensuremath{M_\star}}
\newcommand{\fehstar}{\ensuremath{\feh_\star}}
\newcommand{\lstar}{\ensuremath{L_\star}}

\newcommand{\teffstar}{\ensuremath{T_{\rm eff\star}}}

\newcommand{\loggstar}{\ensuremath{\log{g_{\star}}}}

\newcommand{\mearth}{\ensuremath{M_\earth}}

\newcommand{\rpl}{\ensuremath{R_{\mathrm p}}}
\newcommand{\mpl}{\ensuremath{M_{\mathrm p}}}

\newcommand{\rhopl}{\ensuremath{\rho_{\mathrm p}}}

\newcommand{\arstar}{\ensuremath{a/\rstar}}
\newcommand{\zrstar}{\ensuremath{\zeta/\rstar}}
\newcommand{\rjup}{\ensuremath{R_{\rm J}}}
\newcommand{\mjup}{\ensuremath{M_{\rm J}}}

\newcommand{\reffigl}[1]{Figure~\ref{fig:#1}}
\newcommand{\refsecl}[1]{\mbox{Section \ref{sec:#1}}}

\newcommand{\reftabl}[1]{Table~\ref{tab:#1}}

\newcommand{\flwof}{\mbox{FLWO 1.2 m}}

\newcommand{\hd}[1]{\mbox{HD #1}}

\newcommand{\hatcurhtr}{HTR378-001}                                    
\newcommand{\hatcurCCra}{\ensuremath{14^{\mathrm h}51^{\mathrm m}04.32^{\mathrm s}}}   
\newcommand{\hatcurCCdec}{\ensuremath{+05{\arcdeg}56{\arcmin}50.5{\arcsec}}}          
\newcommand{\hatcurCCtwomass}{2MASS~14510418+0556505}                  
\newcommand{\hatcurCCgsc}{GSC~0333-00351}                              
\newcommand{\hatcurCCtassmv}{12.214}                                   
\newcommand{\hatcurCCtwomassJmag}{\ensuremath{10.626\pm0.026}}         
\newcommand{\hatcurCCtwomassHmag}{\ensuremath{10.249\pm0.023}}         
\newcommand{\hatcurCCtwomassKmag}{\ensuremath{10.109\pm0.021}}         
\newcommand{\hatcurCCesoJKmag}{\ensuremath{0.550\pm0.037}}             
\newcommand{\hatcurLCdip}{\ensuremath{10.6}}                           
\newcommand{\hatcurLCrprstar}{\ensuremath{0.1186\pm0.0031}}            
\newcommand{\hatcurLCbsq}{\ensuremath{0.690_{-0.032}^{+0.029}}}        
\newcommand{\hatcurLCimp}{\ensuremath{0.831_{-0.020}^{+0.017}}}        
\newcommand{\hatcurLCzeta}{\ensuremath{38.20\pm0.80}}                  
\newcommand{\hatcurLCdur}{\ensuremath{0.0705\pm0.0019}}                
\newcommand{\hatcurLCdurhr}{\ensuremath{1.693\pm0.046}}                
\newcommand{\hatcurLCq}{\ensuremath{0.0232\pm0.0006}}                  
\newcommand{\hatcurLCingdur}{\ensuremath{0.0214\pm0.0053}}             
\newcommand{\hatcurLCP}{\ensuremath{3.039586\pm0.000012}}              
\newcommand{\hatcurLCPprec}{\ensuremath{3.039586}}                     
\newcommand{\hatcurLCPshort}{\ensuremath{3.0396}}                      
\newcommand{\hatcurLCT}{\ensuremath{2455186.01879\pm0.00054}}          
\newcommand{\hatcurSMEiteff}{\ensuremath{5350\pm90}}                   
\newcommand{\hatcurSMEizfeh}{\ensuremath{+0.31\pm0.08}}                
\newcommand{\hatcurSMEizfehshort}{\ensuremath{+0.31}}                  
\newcommand{\hatcurSMEilogg}{\ensuremath{4.61\pm0.06}}                 
\newcommand{\hatcurSMEivsin}{\ensuremath{2.7\pm0.5}}                   
\newcommand{\hatcurSMEivmac}{\ensuremath{3.37}}                        
\newcommand{\hatcurSMEivmic}{\ensuremath{0.85}}                        
\newcommand{\hatcurSMEiiteff}{\ensuremath{5300\pm90}}                  
\newcommand{\hatcurSMEiizfeh}{\ensuremath{+0.29\pm0.10}}               
\newcommand{\hatcurSMEiizfehshort}{\ensuremath{+0.29}}                 
\newcommand{\hatcurSMEiilogg}{\ensuremath{4.50\pm0.06}}                
\newcommand{\hatcurSMEiivsin}{\ensuremath{0.4\pm0.4}}                  
\newcommand{\hatcurSMEiivmac}{\ensuremath{3.29}}                       
\newcommand{\hatcurSMEiivmic}{\ensuremath{0.85}}                       
\newcommand{\hatcurTRESteff}{\ensuremath{5250\pm100}}                  
\newcommand{\hatcurTRESlogg}{\ensuremath{4.75\pm0.25}}                 
\newcommand{\hatcurTRESvsini}{\ensuremath{1.5\pm1.0}}                  
\newcommand{\hatcurTRESgamma}{\ensuremath{-15.765\pm0.51}}             
\newcommand{\hatcurTRESrvrms}{\ensuremath{0.05}}                       
\newcommand{\hatcurLBii}{\ensuremath{0.3627}}                          
\newcommand{\hatcurLBiii}{\ensuremath{0.2816}}                         
\newcommand{\hatcurISOm}{\ensuremath{0.94\pm0.04}}                     
\newcommand{\hatcurISOmlong}{\ensuremath{0.945\pm0.035}}               
\newcommand{\hatcurISOr}{\ensuremath{0.90_{-0.04}^{+0.05}}}            
\newcommand{\hatcurISOrlong}{\ensuremath{0.898_{-0.039}^{+0.054}}}     
\newcommand{\hatcurISOlogg}{\ensuremath{4.51\pm0.04}}                  
\newcommand{\hatcurISOlum}{\ensuremath{0.57_{-0.07}^{+0.09}}}          
\newcommand{\hatcurISOmv}{\ensuremath{5.55\pm0.17}}                    
\newcommand{\hatcurISOage}{\ensuremath{4.4_{-2.6}^{+3.8}}}             
\newcommand{\hatcurISOMK}{\ensuremath{3.62\pm0.12}}                    
\newcommand{\hatcurISOJK}{\ensuremath{0.50\pm0.04}}                    
\newcommand{\hatcurISOJKred}{\ensuremath{0.514\pm0.04}}                
\newcommand{\hatcurISOspec}{G8}                                        
\newcommand{\hatcurRVK}{\ensuremath{96.1\pm4.5}}                       
\newcommand{\hatcurRVk}{\ensuremath{0.036\pm0.031}}                    
\newcommand{\hatcurRVh}{\ensuremath{0.066\pm0.048}}                    
\newcommand{\hatcurRVjitter}{\ensuremath{6.3}}                         
\newcommand{\hatcurRVeccen}{\ensuremath{0.078\pm0.047}}                
\newcommand{\hatcurRVomega}{\ensuremath{63\pm64}}                      
\newcommand{\hatcurPPi}{\ensuremath{84.7_{-0.7}^{+0.4}}}               
\newcommand{\hatcurPPlogg}{\ensuremath{3.18\pm0.05}}                   
\newcommand{\hatcurPPar}{\ensuremath{9.65_{-0.54}^{+0.40}}}            
\newcommand{\hatcurPParel}{\ensuremath{0.0403\pm0.0005}}               
\newcommand{\hatcurPPrho}{\ensuremath{0.73\pm0.13}}                    
\newcommand{\hatcurPPmshort}{\ensuremath{0.66}}                        
\newcommand{\hatcurPPmlong}{\ensuremath{0.660\pm0.033}}                
\newcommand{\hatcurPPrshort}{\ensuremath{1.04}}                        
\newcommand{\hatcurPPrlong}{\ensuremath{1.038_{-0.058}^{+0.077}}}      
\newcommand{\hatcurPPmrcorr}{\ensuremath{0.310}}                        
\newcommand{\hatcurPPteff}{\ensuremath{1207\pm41}}                     
\newcommand{\hatcurPPtheta}{\ensuremath{0.054\pm0.004}}                
\newcommand{\hatcurPPfluxperi}{\ensuremath{5.61_{-0.87}^{+1.64}}}      
\newcommand{\hatcurPPfluxperidim}{\ensuremath{8}}                      
\newcommand{\hatcurPPfluxap}{\ensuremath{4.09\pm0.48}}                 
\newcommand{\hatcurPPfluxapdim}{\ensuremath{8}}                        
\newcommand{\hatcurPPfluxavg}{\ensuremath{4.79_{-0.56}^{+0.78}}}       
\newcommand{\hatcurPPfluxavgdim}{\ensuremath{8}}                       
\newcommand{\hatcurXsecondary}{\ensuremath{2455187.608\pm0.060}}       
\newcommand{\hatcurXsecdur}{\ensuremath{0.0739\pm0.0061}}              
\newcommand{\hatcurXsecingdur}{\ensuremath{0.0368\pm0.0066}}           
\newcommand{\hatcurXdist}{\ensuremath{204\pm14}}                       

\newcommand{\hatcur}{HAT-P-27}
\newcommand{\hatcurb}{\hatcur\lowercase{b}}
\newcommand{\hatcurCCtassvi}{\ensuremath{0.527\pm0.12}}                  
\newcommand{\hatcurSMEversion}{ii}                                       
\newcommand{\hatcurSMEteff}{\ifthenelse{\equal{\hatcurSMEversion}{i}}{\hatcurSMEiteff}{\hatcurSMEiiteff}}
\newcommand{\hatcurSMEzfeh}{\ifthenelse{\equal{\hatcurSMEversion}{i}}{\hatcurSMEizfeh}{\hatcurSMEiizfeh}}
\newcommand{\hatcurSMEzfehshort}{\ifthenelse{\equal{\hatcurSMEversion}{i}}{\hatcurSMEizfehshort}{\hatcurSMEiizfehshort}}
\newcommand{\hatcurSMElogg}{\ifthenelse{\equal{\hatcurSMEversion}{i}}{\hatcurSMEilogg}{\hatcurSMEiilogg}}
\newcommand{\hatcurSMEvsin}{\ifthenelse{\equal{\hatcurSMEversion}{i}}{\hatcurSMEivsin}{\hatcurSMEiivsin}}
\newcommand{\hatcurSMEvmac}{\ifthenelse{\equal{\hatcurSMEversion}{i}}{\hatcurSMEivmac}{\hatcurSMEiivmac}}
\newcommand{\hatcurSMEvmic}{\ifthenelse{\equal{\hatcurSMEversion}{i}}{\hatcurSMEivmic}{\hatcurSMEiivmic}}

\newcommand{\hatcurisoshort}{YY}

\newcommand{\hatcurisocite}{yi:2001}
\newcommand{\hatcurlumind}{\arstar}
\newcommand{\hatcurjhkfilset}{ESO}

\shortauthors{B\'eky et al.}
\shorttitle{\hatcurb}
    \newcommand{\titledag}{$\dagger$}

\begin{document}

\title{\hatcurb: A Hot Jupiter Transiting a G Star
    on a 3 day orbit\altaffilmark{\titledag}}

\author{
	B.~B\'eky\altaffilmark{1,2},
	G.~\'A.~Bakos\altaffilmark{1},
	J.~Hartman\altaffilmark{1},
	G.~Torres\altaffilmark{1},
	D.~W.~Latham\altaffilmark{1},
	A.~Jord\'an\altaffilmark{3,1},
    P.~Arriagada\altaffilmark{3},
	D.~Bayliss\altaffilmark{4},
	L.~L.~Kiss\altaffilmark{5,6},
	G\'eza Kov\'acs\altaffilmark{5},
	S.~N.~Quinn\altaffilmark{1},
	G.~W.~Marcy\altaffilmark{7},
	A.~W.~Howard\altaffilmark{7},
	D.~A.~Fischer\altaffilmark{8},
	J.~A.~Johnson\altaffilmark{9},
	G.~A.~Esquerdo\altaffilmark{1},
	R.~W.~Noyes\altaffilmark{1},
    L.~A.~Buchhave\altaffilmark{1,10}
	D.~D.~Sasselov\altaffilmark{1},
    R.~P.~Stefanik\altaffilmark{1},
	G.~Perumpilly\altaffilmark{1,11},
	J.~L\'az\'ar\altaffilmark{12},
	I.~Papp\altaffilmark{12},
	P.~S\'ari\altaffilmark{12}
}

\altaffiltext{1}
    {Harvard--Smithsonian Center for Astrophysics, Cambridge, MA}

\altaffiltext{2}
    {email: \texttt{bbeky@cfa.harvard.edu}}

\altaffiltext{3}
    {Departamento de Astronom\'\i{}a y Astrof\'\i{}sica,
    Pontificia Universidad Cat\'olica de Chile, Santiago, Chile}

\altaffiltext{4}
    {Research School of Astronomy and Astrophysics, The Australian
    National University, Weston Creek, ACT, Australia}

\altaffiltext{5}
    {Konkoly Observatory, Budapest, Hungary}

\altaffiltext{6}
    {Sydney Institute for Astronomy, School of Physics, University of Sydney, Sydney, Australia}

\altaffiltext{7}
    {Department of Astronomy, University of California, Berkeley, CA}

\altaffiltext{8}
    {Department of Astronomy, Yale University, New Haven, CT}

\altaffiltext{9}
    {Astronomy Department, California Institute of Technology, Pasadena, CA}

\altaffiltext{10}
    {Niels Bohr Institute, Copenhagen University, Denmark}

\altaffiltext{11}
    {Department of Physics, University of South Dakota, Vermillion, SD}

\altaffiltext{12}
    {Hungarian Astronomical Association, Budapest, Hungary}

\altaffiltext{$\dagger$}{
	Based in part on observations obtained at the W.~M.~Keck
	Observatory, which is operated by the University of California and
	the California Institute of Technology. Keck time has been
	granted by NOAO (A201Hr) and NASA (N018Hr, N167Hr).
}

\begin{abstract}

\setcounter{footnote}{10}

We report the discovery of \hatcurb{}, an exoplanet transiting
the moderately bright \hatcurISOspec\ dwarf star \hatcurCCgsc{}
($V=\hatcurCCtassmv$). The orbital period is \hatcurLCP{} d,
the reference epoch of transit is \hatcurLCT{} (BJD),
and the transit duration is \hatcurLCdur{} d.
The host star with its effective temperature \hatcurSMEteff{} K is
somewhat cooler than the Sun,
and is more metal-rich with a metallicity of \hatcurSMEzfeh{}.
Its mass is \hatcurISOm{} \msun{}
and radius is \hatcurISOr{} \rsun{}.
For the planetary companion we determine
a mass of \hatcurPPmlong{} \mjup{}
and radius of \hatcurPPrlong{} \rjup.
For the 30 known transiting exoplanets between 0.3 \mjup{} and 0.8 \mjup,
a negative correlation between host star metallicity and planetary radius,
and an additional dependence of planetary radius on equilibrium temperature
are confirmed at a high level of statistical significance.
\setcounter{footnote}{0}
\end{abstract}


\section{Introduction}
\label{sec:introduction}

Studying exoplanets is vital for understanding our own Solar
System, particularly its formation. The sample of more than 500
confirmed exoplanets\footnote{According to
\texttt{http://www.exoplanet.eu/catalog-all.php}.}
so far enables us, for example, to
test accretion and migration theories \citep{ida:2008},
study tidal interactions \citep{mardling:2007},
examine atmospheric structures \citep{fortney:2010},
and investigate correlations between the existence of
planetary companions and the host star's metallicity \citep{ida:2004},
and between the mass of close-in planets
and the spectral type of their host star \citep{ida:2005}.

Among these planets, transiting ones are
of special significance, because
the transit parameters yield planetary mass and radius estimates.
They also provide a means to determine some of the
stellar parameters more accurately than is possible with spectroscopy
alone, such as the stellar surface gravity. The more than 100 transiting
exoplanets confirmed to date provide a sample large enough
to draw meaningful conclusions about the planetary parameters
that could not be determined by radial velocity (RV)
data alone; for example,
the correlation between stellar chromospheric activity and
planetary surface gravity \citep{hartman:2010}, or
the correlation of planetary parameters with host star metallicity
and planetary equilibrium temperature,
as described in \refsecl{discussion}.

The Hungarian-made Automated Telescope Network
\citep[HATNet;][]{bakos:2011} is a system of fully automated
wide-field small telescopes designed to detect the small
photometric dips when exoplanets transit their host stars.
Since 2006, HATNet has announced and 
published 26 planetary systems with 28 planets in total, 26 of which
transit their host stars.

Here we report the detection of our
twenty-seventh transiting exoplanet, named \hatcurb{}, around
the relatively bright \hatcurISOspec{} dwarf known as
\hatcurCCgsc{}.
This planet is a textbook example of a transiting exoplanet with its
radius \hatcurPPrshort{} \rjup{} and orbital period \hatcurLCPshort{} d
being close to the median values for currently known transiting exoplanets,
and with its mass of \hatcurPPmshort{} \mjup{}
being not much less than the median mass of transiting exoplanets.

In \refsecl{obs} we present the photometric detection of the transit,
along with photometric and spectroscopic follow-up observations of
the host star \hatcur{}.
In \refsecl{analysis}, we describe the analysis of the data,
first ruling out false positive scenarios,
then determining parameters of the host star,
and finally performing a global fit for all observational data.
We conclude the paper by discussing \hatcurb{} in the context of other
known transiting exoplanets and investigating correlations of planetary
parameters with host star metallicity and equilibrium temperature
in \refsecl{discussion}.

\section{Observations}
\label{sec:obs}

\subsection{Photometric detection}
\label{sec:detection}

Transits of \hatcurb{} were detected in two HATNet fields
containing its host star \hatcurCCgsc{}, also known as
\hatcurCCtwomass{}; $\alpha = \hatcurCCra$, $\delta = \hatcurCCdec$,
J2000, V=\hatcurCCtassmv{} \citep{droege:2006}; hereafter \hatcur{}.
These fields were observed in Sloan \band {r} on a nightly basis,
weather conditions permitting, from 2009 January to August,
with the HAT-6 and HAT-10 instruments on Mount Hopkins,
and with the HAT-9 instrument on Mauna Kea.
In total, we took 10\,600 science frames
with 5 minute exposure time and 5.5 minute cadence.
For approximately 1200 of the images,
photometric measurements of individual stars had significant error,
therefore these frames were rejected.
Each image contains approximately 20\,000 stars down to $r \approx 14.5$.
For the brightest stars, the per-point photometric precision
is approximately 4.5 mmag.

\begin{figure}[!ht]
\plotone{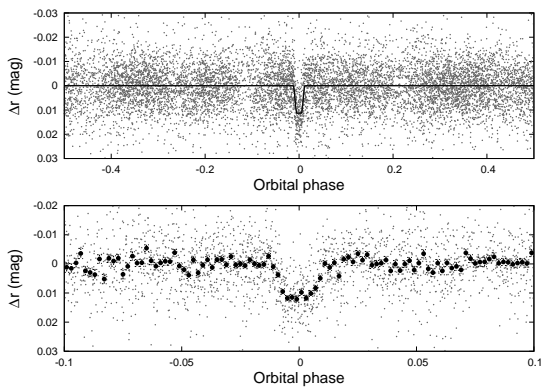}
\caption{
    {\em Top panel:} Unbinned photometric data of \hatcur{} consisting
        of 9\,400 Sloan \band{r} 5.5 minute cadence measurements
        obtained with HATNet telescopes, folded with period $P =
        \hatcurLCPprec$ d.  A simple transit curve fit to the data
        points is displayed with a solid line.  See \refsecl{globmod} for
        details.
    {\em Bottom panel:} A close-up view of the transit. Small gray
        dots are the same folded data as above; large black dots
        show the light curve binned in phase using a
        bin size of 0.002.
\label{fig:hatnet}}
\end{figure}

Calibration, astrometry, aperture photometry, External Parameter
Decorrelation (EPD), the Trend Filtering Algorithm (TFA) and the Box-fitting
Least Squares method were applied to the data as described in
\citet{bakos:2010a}.
We detected a transit signature
in the \lc{} of \hatcur, with a 
signal depth of $\hatcurLCdip$ mmag and period of
$P=\hatcurLCPshort$ days.  This presumed transit
has relative first-to-last-contact duration $q =
\hatcurLCq$, corresponding to total duration
\hatcurLCdur~days (\hatcurLCdurhr~hours).
The folded \lc{} is presented in \reffigl{hatnet}.

\subsection{Reconnaissance Spectroscopy}
\label{sec:recspec}

Reconnaissance spectra \citep{latham:2009} of \hatcur{} were taken using three
facilities: the Tillinghast Reflector Echelle Spectrograph
\citep[TRES;][]{furesz:2008} on the 1.5 m Tillinghast Reflector at
FLWO, the echelle spectrograph on the 2.5 m du Pont telescope at Las
Campanas Observatory (LCO) in Chile, and the echelle spectrograph on
the Australian National University (ANU) 2.3 m telescope
at Siding Spring Observatory (SSO) in Australia. 
We gathered two spectra of \hatcur{} with TRES in
2009 July and 2010 February,
two spectra with the du Pont telescope
in 2009 July, and fourteen spectra with the ANU
2.3 m telescope in 2009 July. The exact dates and the results of these
observations are summarized in \reftabl{reconspecobs}.

\begin{deluxetable*}{llccccr}
\tablewidth{0pc}
\tabletypesize{\scriptsize}
\tablecaption{
    Summary of reconnaissance spectroscopy observations of \hatcur{}
    \label{tab:reconspecobs}
}
\tablehead{
    \multicolumn{1}{c}{Instrument}          &
    \multicolumn{1}{c}{Date}                &
    \multicolumn{1}{c}{Number of}           &
    \multicolumn{1}{c}{$\teffstar$}         &
    \multicolumn{1}{c}{$\loggstar$}         &
    \multicolumn{1}{c}{$\vsini$}            &
    \multicolumn{1}{c}{$\gamma_{\rm RV}$\tablenotemark{a}} \\
    &
    &
    \multicolumn{1}{c}{Spectra}             &
    \multicolumn{1}{c}{(K)}                 &
    \multicolumn{1}{c}{(cgs)}               &
    \multicolumn{1}{c}{(\kms)}              &
    \multicolumn{1}{c}{(\kms)}
}
\startdata
TRES      & 2009 Jul 05 & 1 & $5250$  & $4.5$   & $2$     & $-15.75$ \\
du Pont   & 2009 Jul 10 & 1 & $5250$  & $4.0$   & $0$     & $-16.03$ \\
du Pont   & 2009 Jul 11 & 1 & $5000$  & $3.5$   & $0$     & $-18.03$ \\
ANU 2.3 m & 2009 Jul 16 & 5 & \nodata & \nodata & \nodata & $-20.58$ \\
ANU 2.3 m & 2009 Jul 17 & 6 & \nodata & \nodata & \nodata & $-21.32$ \\
ANU 2.3 m & 2009 Jul 18 & 3 & \nodata & \nodata & \nodata & $-20.61$ \\
TRES      & 2010 Feb\,13& 1 & $5250$  & $5.0$   & $1$     & $-15.78$ \\[-2ex]
\enddata 
\tablenotetext{a}{
    The mean heliocentric RV of the target. Systematic differences
    between the velocities from different instruments are consistent
    with the velocity zero-point uncertainties.
}
\end{deluxetable*}

Following \citet{quinn:2010} and \citet{buchhave:2010},
we calculated the mean radial velocity, effective temperature,
surface gravity, and projected rotational velocity of the host star,
based on spectra taken by TRES.
The inferred radial velocity RMS residual of \hatcurTRESrvrms{} \kms{}
is consistent with no detectable RV
variation within the precision of the measurements.
We established the following atmospheric parameters for \hatcur{}:
effective temperature $\teffstar =
\hatcurTRESteff$ K, surface gravity $\loggstar = \hatcurTRESlogg$
(cgs), and projected rotational velocity $\vsini =
\hatcurTRESvsini$ \kms, indicating a
\hatcurISOspec\ dwarf. The mean heliocentric RV of \hatcur{} after
subtracting the gravitational redshift of the Sun is
$\gamma_{\rm RV} = \hatcurTRESgamma$ \kms.

Because this is the first time we used the du Pont 2.5 m and ANU 2.3 m
telescopes for reconnaissance spectroscopy of HATNet targets, we
briefly describe the instruments and our data reduction procedure.

The spectrograph on the du Pont telescope was used with a $4\arcsec$
long and $1.5\arcsec$ wide slit. The obtained spectra have
wavelength coverage $\approx$ 3700--7000 \AA{} at a resolution of
$\lambda/\Delta\lambda \approx 26\,000$. During the first observation
the seeing ranged between 2--3\arcsec\ and we used an exposure time of
1200 s, which provided $\sim 3000$ electrons per resolution element at 
the wavelength of 5187 \AA\@. The seeing during the second observation
was $\approx
1.8\arcsec$ and we used an exposure time of 150 s to obtain a lower
S/N spectrum sufficient to detect a velocity variation of several
\kms. We obtained a ThAr lamp spectrum before and after each
observation to use in determining the wavelength solution. We used the
CCDPROC package of IRAF\footnote{IRAF is distributed by the National
  Optical Astronomical Observatory, which is operated by the
  Association of Universities for Research in Astronomy (AURA) under
  cooperative agreement with the National Science Foundation.} to
perform overscan correction and flat-fielding of the images,
and the ECHELLE package to extract the spectra
and to determine and apply the dispersion corrections.

The extracted du Pont spectra were then cross-correlated against a
library of synthetic stellar spectra to estimate the effective
temperature, surface gravity, projected rotational velocity, and
radial velocity of the star. We followed a procedure similar to that
described by \citet{torres:2002}, using the same synthetic templates,
but broadened to the resolution of the du Pont echelle. These
templates, which were generated for the CfA Digital Speedometer
\citep{latham:1992}, only cover a wavelength range of
5150--5360 \AA{}, so we restricted our analysis to a single order of
the spectrum covering a similar range.

Spectra were also taken using the echelle spectrograph on the ANU 2.3 m
telescope.  The echelle was used in a standard configuration with a
$1.8\arcsec$ wide slit and $300\;\mathrm{mm}^{-1}$ cross-disperser setting
of $5^{\circ}50'$, which delivered 27 full orders between 
3900--6720 \AA{} with a nominal spectral resolution of
$\lambda/\Delta\lambda \approx 23\,000$.  The CCD is a 2K$\times$2K
format with $13.5\;\mu\mathrm m \times 13.5\;\mu\mathrm m$ pixels.
The gain was two electrons per ADU resulting in a read noise
of approximately 2.3 ADU for each pixel.  The spectra were binned by
two along the spatial direction.
A total of fourteen 1200 s exposures were taken.  
The seeing ranged from $2\arcsec$ to $3\arcsec$.  
The signal-to-noise on a single pixel was between 5 and 10 for each
individual exposure.  ThAr lamp calibration exposures were taken every
hour for wavelength calibration.  A high signal-to-noise exposure was
also taken of the bright radial velocity standard star \hd{223311}.

Spectra were reduced using tasks in the IRAF packages CCDPROC and
ECHELLE.  The spectra were cross-correlated against the radial velocity
standard star \hd{223311} using the IRAF task FXCOR in the RV package.
We used at least 20, typically 25 of the 27 orders for the
cross-correlation, rejecting the bluest orders for many exposures where
the signal-to-noise was too low.  Each spectral order was treated
separately and the mean of the velocities from the individual orders
was calculated. Their standard deviations were less than $0.65\;\kms$
for each exposure.

Each night, the exposures were taken within a two hour interval, much
shorter than the orbital period of the presumed companion.
For detecting large radial velocity variations to rule out the
possibility of an eclipsing binary, we consider the mean radial
velocities per night. The standard deviations between exposures were
less than $0.75\;\kms$ for each night.
Stellar parameters could have been estimated only with large
uncertainty based on data with such low signal-to-noise. Therefore
these parameters are not calculated from the
ANU 2.3 m observations. 

The results of the observations taken with these three telescopes
are listed in \reftabl{reconspecobs}. Note that for each telescope,
the radial velocity measurement uncertainty is much higher than the
radial velocity variations of the Sun due to Solar System bodies.
Therefore we only calculated heliocentric radial velocities of the
target. For the more accurate measurements described in the next section,
we will use barycentric radial velocities instead.
This accuracy, however, is enough to rule out the case of an eclipsing
binary star with great certainty. The largest RV variation {\em within an
instrument} was only 2 \kms, much less than the orbital speed in a 
typical binary system. Note that the zero-point shift between
instruments is as large as $\sim5$ \kms{}, due to the different
methods used for data reduction.
Also, all eighteen spectra were single-lined
and spectral lines were symmetric, providing no evidence for
additional stars in the system up to the precision of the measurements.

\subsection{High resolution, high S/N spectroscopy}
\label{sec:hispec}

We acquired high-resolution, high-S/N spectra of \hatcur{} using the
HIRES instrument \citep{vogt:1994} on the Keck~I telescope located on
Mauna Kea, Hawaii, between 2009 December and 2010 June.
The spectrometer was configured with the $0.86\arcsec$ wide slit,
yielding a resolving power of 
$\lambda/\Delta\lambda \approx 55\,000$ over the wavelength range
of $\approx$ 3800--8000 \AA\@.

\begin{figure} [ht]
\plotone{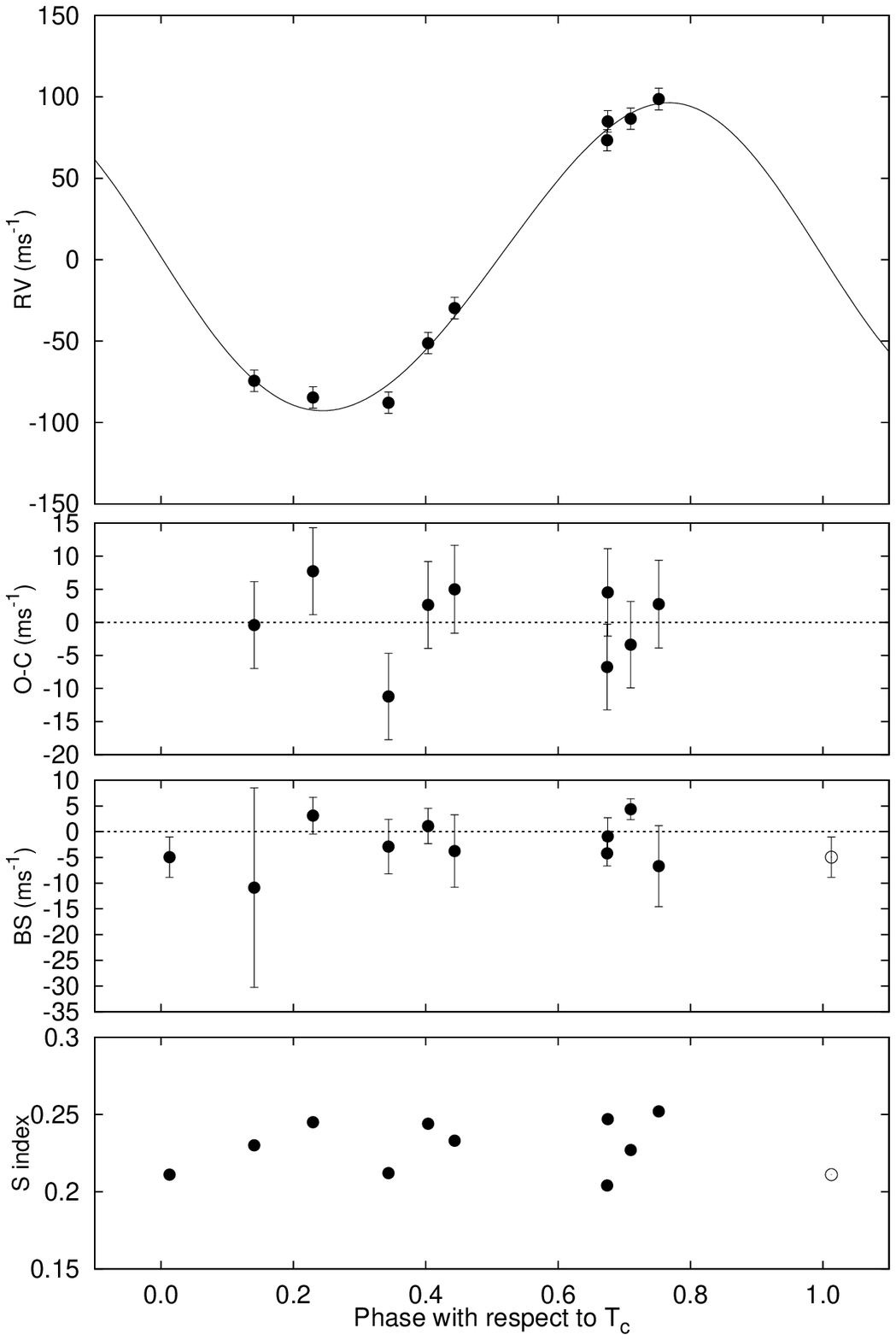}
\caption{
	{\em Top panel:} Keck/HIRES RV measurements for \hbox{\hatcur{}}
    shown as a function of orbital phase, along with our best-fit model
    (see \reftabl{planetparam}).  Zero phase corresponds to the time of
    mid-transit.  The center-of-mass velocity has been subtracted.
	{\em Second panel:} Velocity $O\!-\!C$ residuals from the best fit. 
    The error bars include a component from astrophysical/instrumental
    jitter ($\hatcurRVjitter$ \ms) added in quadrature to the formal
    errors (see \refsecl{globmod}).
	{\em Third panel:} Bisector spans (BS), with the mean value
        subtracted, and corrected for sky contamination.  The
        measurement from the template spectrum is included (see
        \refsecl{blend}).
	{\em Bottom panel:} Chromospheric activity index $S$.  Again, the
        measurement from the template spectrum is included.
	{\em Note:} Panels have different vertical scales. The data point
    replotted in the second period is represented by an open symbol.
\label{fig:rvbis}}
\end{figure}

Nine exposures were taken using an $\mathrm{I}_2$ gas cell
\citep[see][]{marcy:1992},
and a single template exposure was obtained without the absorption cell.
We followed \cite{butler:1996} to establish RVs in the Solar System
barycentric frame.
We also calculated the $S$ index for each spectrum
\citep[following][]{isaacson:2010}.
The resulting values and their uncertainties are listed in \reftabl{rvs}.
They are plotted period-folded in \reffigl{rvbis}, together with the
fit established in \refsecl{analysis}.

The effective temperature established later in \refsecl{stelparam}
compared to Figure 4~of \citet{valenti:2005} implies $B-V=0.800$ for the star.
This can in turn be used in the formula given in \citet{noyes:1984}
together with the median $S_\mathrm{HK}$ of 0.231 to conclude
$\log R'_\mathrm{HK}=-4.785$.
This activity value is consistent with the RV jitter value
\hatcurRVjitter{} \ms{} established in \refsecl{globmod},
according to the observations given by \citet{wright:2005}.
Based on Figure 9 in \citet{isaacson:2010}, this value qualifies
\hatcur{} as chromospherically active relative to California
Planet Search targets of the same spectral class.
The
activity index does not correlate significantly
with orbital phase.

\begin{deluxetable*}{lrrrrrr}
\tablewidth{0pc}
\tablecaption{
	Relative radial velocities, bisector spans, and activity index
	measurements of \hatcur{}
	\label{tab:rvs}
}
\tablehead{
	\colhead{BJD (UTC)} & 
	\colhead{RV\tablenotemark{a}} & 
	\colhead{\ensuremath{\sigma_{\rm RV}}\tablenotemark{b}} & 
	\colhead{BS\tablenotemark{c}} & 
	\colhead{\ensuremath{\sigma_{\rm BS}}} & 
	\colhead{$S$\tablenotemark{d}} & 
	\colhead{Phase}\\
	\colhead{\hbox{($2\,400\,000+$)}} & 
	\colhead{(\ms)} & 
	\colhead{(\ms)} &
	\colhead{(\ms)} &
    \colhead{(\ms)} &
	\colhead{} &
	\colhead{}
}
\startdata
$ 55192.13748 $ & \nodata {\ } & \nodata      & $    -4.96 $ & $     3.92 $ & $    0.211 $ & $   0.013 $\\
$ 55193.14273 $ & $   -87.86 $ & $     1.73 $ & $    -2.89 $ & $     5.30 $ & $    0.212 $ & $   0.344 $\\
$ 55194.14692 $ & $    73.31 $ & $     1.51 $ & $    -4.20 $ & $     2.47 $ & $    0.204 $ & $   0.674 $\\
$ 55252.00750 $ & $    86.53 $ & $     1.75 $ & $     4.37 $ & $     2.03 $ & $    0.227 $ & $   0.710 $\\
$ 55257.15592 $ & $   -51.27 $ & $     1.84 $ & $     1.11 $ & $     3.44 $ & $    0.244 $ & $   0.404 $\\
$ 55261.02101 $ & $    84.94 $ & $     2.02 $ & $    -0.91 $ & $     3.63 $ & $    0.247 $ & $   0.675 $\\
$ 55290.06295 $ & $   -84.71 $ & $     1.89 $ & $     3.14 $ & $     3.57 $ & $    0.245 $ & $   0.230 $\\
$ 55312.92780 $ & $    98.63 $ & $     2.08 $ & $    -6.69 $ & $     7.87 $ & $    0.252 $ & $   0.752 $\\
$ 55374.90154 $ & $   -74.42 $ & $     1.85 $ & $   -10.88 $ & $    19.39 $ & $    0.230 $ & $   0.141 $\\
$ 55375.82176 $ & $   -29.80 $ & $     2.10 $ & $    -3.76 $ & $     7.04 $ & $    0.233 $ & $   0.444 $\\[-2ex]
\enddata
\tablenotetext{a}{
	The zero-point of these velocities is arbitrary. An overall offset
    $\gamma_{\rm rel}$ fitted to these velocities in \refsecl{globmod}
    has {\em not} been subtracted.
}
\tablenotetext{b}{
    Internal errors excluding the component of
    astrophysical/instrumental jitter considered in \refsecl{globmod}.
}
\tablenotetext{c}{
    The bisector spans have been corrected for sky contamination
    following \citet{hartman:2011}.
}
\tablenotetext{d}{
	Relative chromospheric activity index, calibrated to the
	scale of \citet{vaughan:1978}.
}
\tablecomments{
    For the iodine-free template exposures there is no RV
    measurement, but the BS and $S$ index can still be determined.
}
\end{deluxetable*}

\subsection{Photometric follow-up observations}
\label{sec:phot}

\begin{figure}[!ht]
\plotone{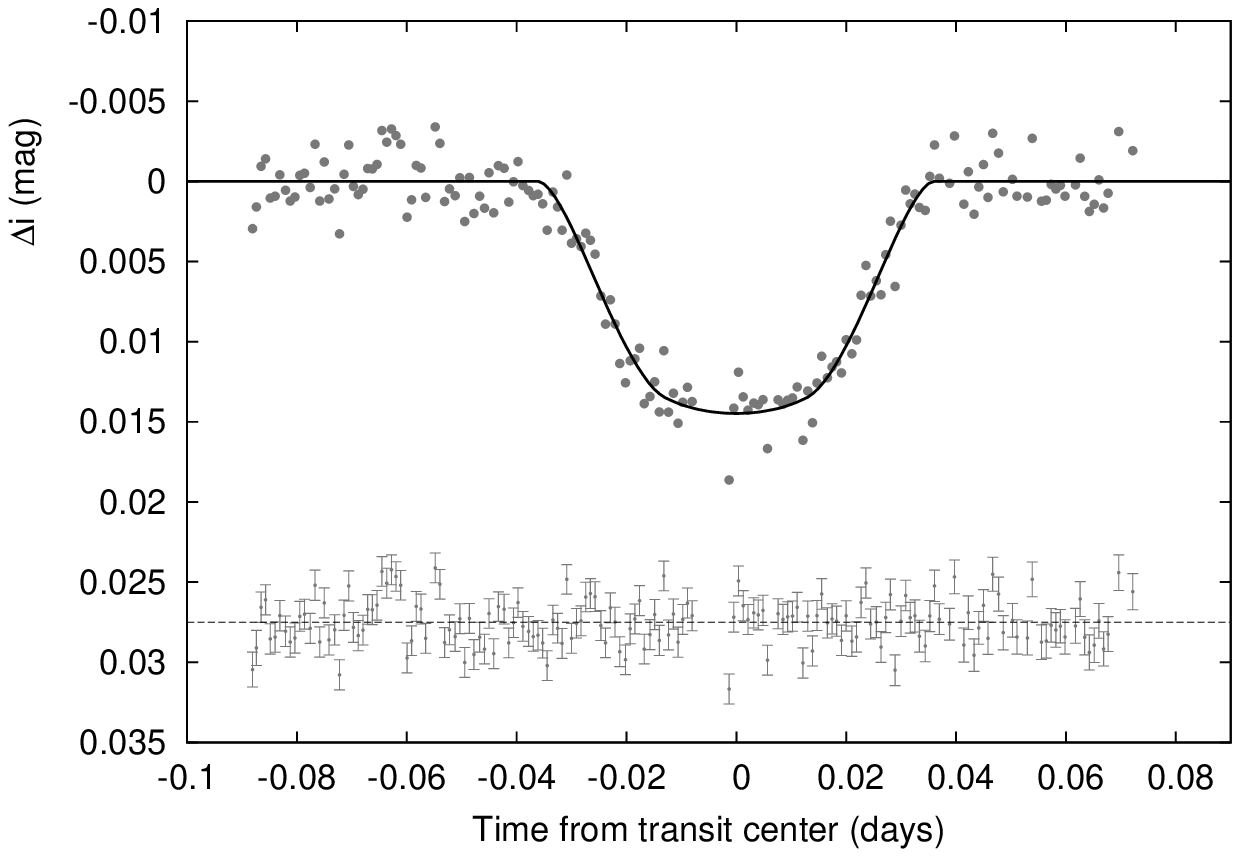}
\caption{
	Unbinned Sloan \band{i} transit light curve, acquired
	with KeplerCam on the \flwof{} telescope on 2010 March 2.
	The light curve has been EPD and TFA-processed.
    Our best fit from the global modeling is shown by the solid line.
    See \refsecl{globmod} for details.  Residuals from the
    fit are displayed at the bottom.  The error bars represent the
    photon and background shot noise, plus the readout noise.
\label{fig:lc}}
\end{figure}

A high-precision photometric follow-up of a complete transit was
carried out, permitting refined estimates of the light curve parameters
and thus orbital and planetary properties: The transit of
\hatcur{} was observed on the night of 2010 March 2 with the
KeplerCam CCD camera on the \flwof{} telescope.
We acquired 168 science frames in Sloan \band {i}
with 60 second exposure time, 73 second cadence.

Following the procedure described by \citet{bakos:2010a},
these images were first calibrated, then astrometry and aperture photometry
was performed to arrive at \lcs, which were finally cleaned of trends
using EPD and TFA, carried out simultaneously with the global modeling
described in \refsecl{globmod}.
The result is shown in \reffigl{lc}, along with the best-fit
transit \lc{}; the individual measurements are
reported in \reftabl{phfu}.

\begin{deluxetable*}{lrrrr}
\tablewidth{0pc}
\tablecaption{High-precision differential photometry of \hatcur\label{tab:phfu}}
\tablehead{
	\colhead{BJD (UTC)} & 
	\colhead{Mag\tablenotemark{a}} & 
	\colhead{\ensuremath{\sigma_{\rm Mag}}} &
	\colhead{Mag(orig)\tablenotemark{b}} & 
	\colhead{Filter} \\
	\colhead{\hbox{($2\,400\,000+$)}} & 
	\colhead{} & 
	\colhead{} &
	\colhead{} & 
	\colhead{}
}
\startdata
$ 55258.88089 $ & $   0.00295 $ & $   0.00109 $ & $  10.86800 $ & $ i$\\
$ 55258.88157 $ & $   0.00160 $ & $   0.00109 $ & $  10.86810 $ & $ i$\\
$ 55258.88242 $ & $  -0.00094 $ & $   0.00095 $ & $  10.86430 $ & $ i$\\
$ 55258.88327 $ & $  -0.00140 $ & $   0.00094 $ & $  10.86320 $ & $ i$\\
$ 55258.88412 $ & $   0.00104 $ & $   0.00095 $ & $  10.86640 $ & $ i$\\
$ 55258.88499 $ & $   0.00093 $ & $   0.00094 $ & $  10.86500 $ & $ i$\\
$ 55258.88584 $ & $  -0.00041 $ & $   0.00095 $ & $  10.86470 $ & $ i$\\
$ 55258.88689 $ & $   0.00056 $ & $   0.00095 $ & $  10.86350 $ & $ i$\\
$ 55258.88773 $ & $   0.00123 $ & $   0.00094 $ & $  10.86580 $ & $ i$\\
$ 55258.88859 $ & $   0.00098 $ & $   0.00094 $ & $  10.86540 $ & $ i$\\[-2ex]
\enddata
\tablenotetext{a}{
	The out-of-transit level has been subtracted. These magnitudes have
	been obtained by the EPD and TFA procedures, carried out
	simultaneously with the transit fit.
}
\tablenotetext{b}{
	Raw magnitude values without application of the EPD and TFA
	procedures.
}
\tablecomments{
    This table is available in a machine-readable form in the online
    journal.  A portion is shown here for guidance regarding its form
    and content.
}
\end{deluxetable*}

\section{Analysis}
\label{sec:analysis}

\subsection{Excluding blend scenarios}
\label{sec:blend}

To further exclude possible blends, we perform bisector analysis
the same way as in \S 5 of \cite{bakos:2007}.  A significant scatter
is found, strongly correlated to the presence of moonlight, which we
account for using the method described by \cite{hartman:2011}.  The
bisector spans, corrected for the effect of the moonlight, are shown 
in the third panel of \reffigl{rvbis}.  They do not exhibit
significant correlation with the RV values, and the RMS scatter of the
bisector spans (4.6 \ms{}) is significantly smaller than the RV
amplitude.  These findings rule out a blend scenario with high
certainty, implying that the measured photometric and spectroscopic
features are due to a planet orbiting \hatcur{}.

\subsection{Properties of the parent star}
\label{sec:stelparam}

We first determine spectroscopic parameters of \hatcur{},
which will allow us to calculate stellar mass and radius.
The Spectroscopy Made Easy analysis package \citep[SME;][]{valenti:1996}
is used to establish the effective temperature, metallicity and
stellar surface gravity
based on the Keck/HIRES template spectrum of \hatcur{},
using atomic line data from the database of \cite{valenti:2005}.
After an initial estimate for these parameters,
we perform a Monte Carlo calculation,
relying also on the normalized semi-major axis \arstar{}
inferred from transit \lcs{},
for the reasons described by \citet {bakos:2010b}.
The final values adopted after two iterations are
$\teffstar = \hatcurSMEiiteff$ K, 
$\feh = \hatcurSMEiizfeh$, and 
$\vsini = \hatcurSMEiivsin$ \kms, also listed in \reftabl{stellar}.

\begin{figure}[!ht]
\plotone{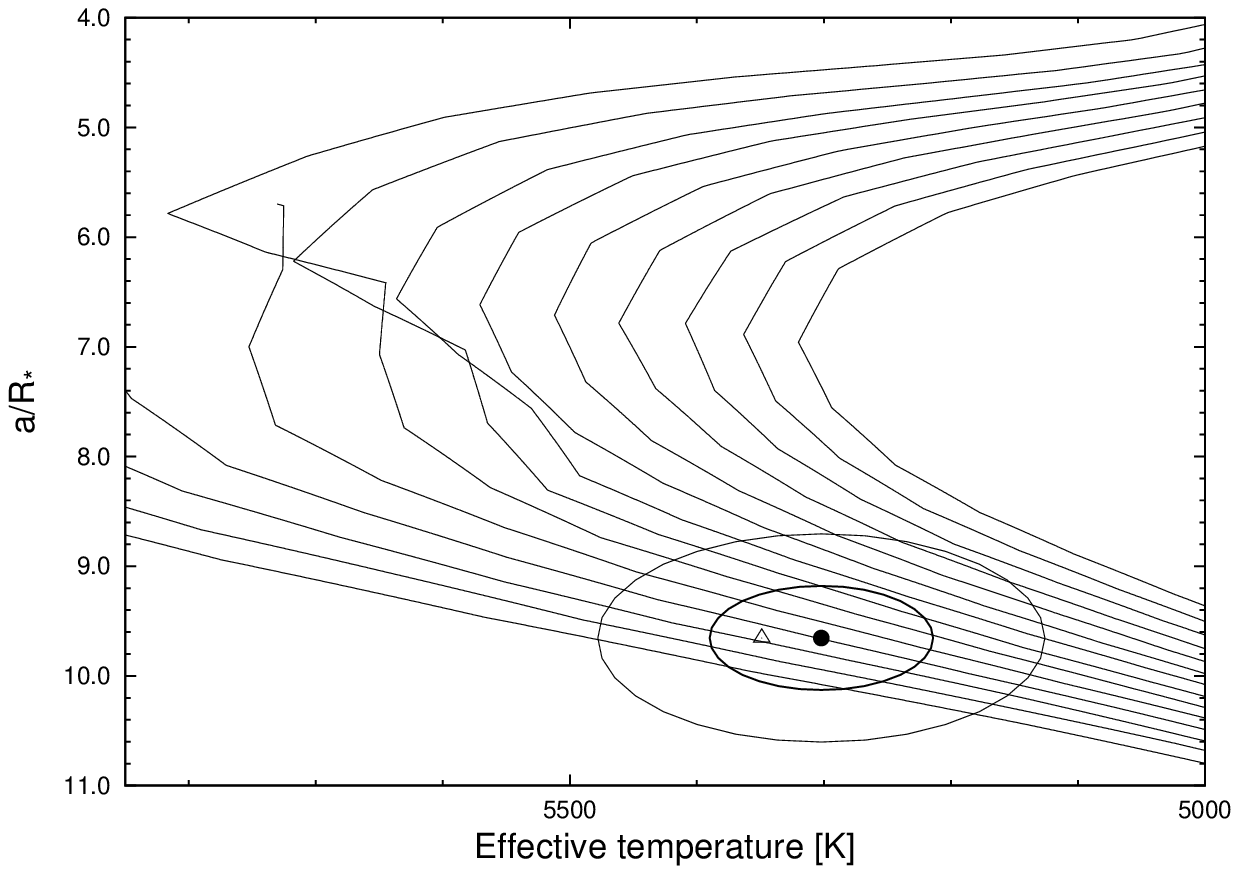}
\caption{
    Model isochrones from \cite{\hatcurisocite} for the measured
    metallicity of \hatcur, \feh = \hatcurSMEiizfehshort, and ages
    running from 1{} Gyr to 14{} Gyr in 1{} Gyr increments,
    left to right.  The adopted values of $\teffstar$ and \arstar{}
    are shown as the solid dot, surrounded by $1\sigma$ and 
    $2\sigma$ confidence ellipsoids.  The
    initial values of \teffstar\ and \arstar\ from the first SME and
    \lc\ analyses are represented with a triangle.
\label{fig:iso}}
\end{figure}

Based on the final spectroscopic parameters the model isochrones yield
a stellar mass \mstar\ = \hatcurISOmlong{} \msun{} and radius
\rstar\ = \hatcurISOrlong{} \rsun{} for \hatcur{},
along with other properties listed at the bottom of \reftabl{stellar}.
These values classify the star as a \hatcurISOspec{} dwarf,
and suggest an age of \hatcurISOage{} Gyr.
The model isochrones of \cite{\hatcurisocite} for a metallicity of
\hatcurSMEiizfehshort{} are plotted in \reffigl{iso}, along with
the best estimate of \arstar{} and \teffstar{} of \hatcur{},
and their $1\sigma$ and $2\sigma$ confidence ellipsoids. 
For comparison, the initial SME result, corresponding to a somewhat
younger state, is also indicated.

The intrinsic absolute magnitude predictions of this model
(given in the ESO photometric system)
can be compared to observations to calculate the distance of \hatcur{}.
For this we use the near-infrared brightness measurements from
the 2MASS Catalogue \citep{skrutskie:2006}:
$J_{\rm 2MASS}=\hatcurCCtwomassJmag$, 
$H_{\rm 2MASS}=\hatcurCCtwomassHmag$ and 
$K_{\rm 2MASS}=\hatcurCCtwomassKmag$.
These values are converted to ESO
\citep{carpenter:2001}, then
compared to the absolute magnitude estimates to calculate the
distance. We account for interstellar dust extinction in the line of
sight using $E(B\!-\!V)=0.036 \pm 0.010$ from the dust map by
\citet{schlegel:1998}\footnote{
\texttt{http://irsa.ipac.caltech.edu/applications/DUST}}
with a conservative uncertainty estimate. This has to be multiplied
by a factor depending on the distance of the star and its Galactic latitude
\citep[see][]{bonifacio:2000}. Assuming diffuse interstellar medium
and no dense clouds along the line of sight, we use the value
$R_V=3.1$, along with the coefficients given in Table 3 in
\citet{cardelli:1989}. These let us calculate extinction parameters
for each band based on the reddening. Finally, comparing extinctions,
absolute magnitude predictions and 2MASS apparent magnitudes for $J$,
$H$ and $K$ bands, we arrive at separate distance estimates. These 
are in good agreement, yielding an average distance of
$\hatcurXdist$ pc. The uncertainty does not account for possible systematics
of the stellar evolution model. Note that this value
is only 1 pc less than the more simple estimate ignoring extinction.
The model predicts an unreddened color index of
$(J-K)_\mathrm{model} = \hatcurISOJK$.
Reddening would change it to an estimated observed value of
$(J-K)_\mathrm{red} = \hatcurISOJKred$ using the above parameters.
This matches the actual observed color index
$(J-K)_\mathrm{ESO} = \hatcurCCesoJKmag$ within $1\sigma$.

\begin{deluxetable}{lrl}
\tablewidth{0pc}
\tabletypesize{\scriptsize}
\tablecaption{
	Stellar parameters for \hatcur{}
	\label{tab:stellar}
}
\tablehead{
	\colhead{~~~~~~~~Parameter~~~~~~~~}	&
	\colhead{Value} &
	\colhead{Source}
}
\startdata
\noalign{\vskip -3pt}
\sidehead{Spectroscopic properties}
~~~~$\teffstar$ (K)\dotfill         &  \hatcurSMEteff       & SME\tablenotemark{a}\\
~~~~$\feh$\dotfill                  &  \hatcurSMEzfeh       & SME                 \\
~~~~$\vsini$ (\kms)\dotfill         &  \hatcurSMEvsin       & SME                 \\
~~~~$\vmac$ (\kms)\dotfill          &  \hatcurSMEvmac       & SME                 \\
~~~~$\vmic$ (\kms)\dotfill          &  \hatcurSMEvmic       & SME                 \\
~~~~$\gamma_{\rm RV}$ (\kms)\dotfill&  \hatcurTRESgamma     & TRES                \\
\sidehead{Photometric properties}
~~~~$V$ (mag)\dotfill               &  \hatcurCCtassmv      & TASS                \\
~~~~$V\!-\!I_C$ (mag)\dotfill       &  \hatcurCCtassvi      & TASS                \\
~~~~$J$ (mag)\dotfill               &  \hatcurCCtwomassJmag & 2MASS           \\
~~~~$H$ (mag)\dotfill               &  \hatcurCCtwomassHmag & 2MASS           \\
~~~~$K_s$ (mag)\dotfill             &  \hatcurCCtwomassKmag & 2MASS           \\
\sidehead{Derived properties}
~~~~$\mstar$ ($\msun$)\dotfill      &  \hatcurISOmlong      & \hatcurisoshort+\hatcurlumind+SME \tablenotemark{b}\\
~~~~$\rstar$ ($\rsun$)\dotfill      &  \hatcurISOrlong      & \hatcurisoshort+\hatcurlumind+SME         \\
~~~~$\loggstar$ (cgs)\dotfill       &  \hatcurISOlogg       & \hatcurisoshort+\hatcurlumind+SME         \\
~~~~$\lstar$ ($\lsun$)\dotfill      &  \hatcurISOlum        & \hatcurisoshort+\hatcurlumind+SME         \\
~~~~$M_V$ (mag)\dotfill             &  \hatcurISOmv         & \hatcurisoshort+\hatcurlumind+SME         \\
~~~~$M_K$ (mag, \hatcurjhkfilset)\dotfill & \hatcurISOMK    & \hatcurisoshort+\hatcurlumind+SME         \\
~~~~Age (Gyr)\dotfill               &  \hatcurISOage        & \hatcurisoshort+\hatcurlumind+SME         \\
~~~~Distance (pc)\dotfill           &  \hatcurXdist         & \hatcurisoshort+\hatcurlumind+SME         \\[-2ex]
\enddata
\tablenotetext{a}{
	SME = Spectroscopy Made Easy package for the analysis of
	high-resolution spectra \citep{valenti:1996}.  These parameters
	rely primarily on SME, but have a small dependence also on the
	iterative analysis incorporating the isochrone search and global
	modeling of the data, as described in the text.
}
\tablenotetext{b}{
	\hatcurisoshort+\hatcurlumind+SME = Based on the \hatcurisoshort\
    isochrones \citep{\hatcurisocite}, \hatcurlumind\ as a luminosity
    indicator, and the SME results.
}
\end{deluxetable}

\subsection{Global modeling of the data}
\label{sec:globmod}

We fit the combined model described by \citet{bakos:2010a} 
to the HATNet photometry, follow-up photometry,
and radial velocity measurements simultaneously.
The eight main parameters describing the model are
the time of the first observed transit center, 
the time of the follow-up transit center, 
the normalized planetary radius $p \equiv \rpl/\rstar$,
the square of the impact parameter $b^2$,
the reciprocal of the half duration of the transit $\zrstar$,
the RV semi-amplitude $K$,
and the Lagrangian elements $k \equiv e\cos\omega$ and $h \equiv e\sin\omega$
(where $\omega$ is the longitude of periastron).

Instrumental parameters of the model include
the HATNet \oot{} magnitude
and the relative zero-point
of the Keck RVs.  The joint fit provides the full {\em a posteriori}
probability distributions of all variables (including \loggstar), which
are used to update the limb-darkening coefficients for another
iteration of the joint fit.  This leads to estimated distributions for
the stellar, \lc, and RV parameters, which are combined to calculate
values for planetary parameters.  These final values are summarized in
\reftabl{planetparam}.

The orbital eccentricity is consistent with zero (using the method of
\citealp{lucy:1971} we find that there is a 25\% conditional
probability of detecting an eccentricity of at least 0.078 given a
circular orbit and an uncertainty of 0.047).
Nevertheless, we stress that throughout 
the global modeling, the orbit was allowed to be eccentric, and all
system parameters and their respective errors 
inherently contain the uncertainty arising from the
floating $k$ and $h$ values.

\begin{deluxetable}{lr}
\tabletypesize{\scriptsize}
\tablecaption{Orbital and planetary parameters\label{tab:planetparam}}
\tablehead{
	\colhead{~~~~~~~~~~~~~~~Parameter~~~~~~~~~~~~~~~} &
	\colhead{Value}
}
\startdata
\noalign{\vskip -3pt}
\sidehead{\Lc{} parameters}
~~~$P$ (days)             \dotfill    & $\hatcurLCP$              \\
~~~$T_c$\tablenotemark{a} (BJD, UTC)    
                          \dotfill    & $\hatcurLCT$              \\
~~~$T_{14}$ (days)
      \tablenotemark{b}   \dotfill    & $\hatcurLCdur$            \\
~~~$T_{12} = T_{34}$ (days)
      \tablenotemark{c}   \dotfill    & $\hatcurLCingdur$         \\
~~~$\arstar$              \dotfill    & $\hatcurPPar$             \\
~~~$\zrstar$              \dotfill    & $\hatcurLCzeta$           \\
~~~$\rpl/\rstar$          \dotfill    & $\hatcurLCrprstar$        \\
~~~$b^2$                  \dotfill    & $\hatcurLCbsq$            \\
~~~$b \equiv a \cos i/\rstar$
                          \dotfill    & $\hatcurLCimp$            \\
~~~$i$ (degree)           \dotfill    & $\hatcurPPi$              \\

\sidehead{Limb-darkening coefficients \tablenotemark{d}}
~~~$a_i$ (linear term)    \dotfill    & $\hatcurLBii$             \\
~~~$b_i$ (quadratic term) \dotfill    & $\hatcurLBiii$            \\

\sidehead{RV parameters}
~~~$K$ (\ms)              \dotfill    & $\hatcurRVK$              \\
~~~$k_{\rm RV}$\tablenotemark{e} 
                          \dotfill    & $\hatcurRVk$              \\
~~~$h_{\rm RV}$\tablenotemark{e}
                          \dotfill    & $\hatcurRVh$              \\
~~~$e$                    \dotfill    & $\hatcurRVeccen$          \\
~~~$\omega$ (degree)      \dotfill    & $\hatcurRVomega$          \\
~~~RV jitter (\ms)
    \tablenotemark{f}     \dotfill    & \hatcurRVjitter           \\

\sidehead{Secondary eclipse parameters}
~~~$T_s$ (BJD)            \dotfill    & $\hatcurXsecondary$       \\
~~~$T_{s,14}$             \dotfill    & $\hatcurXsecdur$          \\
~~~$T_{s,12}$             \dotfill    & $\hatcurXsecingdur$       \\

\sidehead{Planetary parameters}
~~~$\mpl$ ($\mjup$)       \dotfill    & $\hatcurPPmlong$          \\
~~~$\rpl$ ($\rjup$)       \dotfill    & $\hatcurPPrlong$          \\
~~~$C(\mpl,\rpl)$
    \tablenotemark{g}     \dotfill    & $\hatcurPPmrcorr$         \\
~~~$\rhopl$ (\gcmc)       \dotfill    & $\hatcurPPrho$            \\
~~~$\log g_p$ (cgs)       \dotfill    & $\hatcurPPlogg$           \\
~~~$a$ (AU)               \dotfill    & $\hatcurPParel$           \\
~~~$T_{\rm eq}$ (K)       \dotfill    & $\hatcurPPteff$           \\
~~~$\Theta$\tablenotemark{h}\dotfill  & $\hatcurPPtheta$          \\
~~~$F_{per}$ ($10^{\hatcurPPfluxperidim}$ \ergscmsq)
    \tablenotemark{i}     \dotfill    & $\hatcurPPfluxperi$       \\
~~~$F_{ap}$  ($10^{\hatcurPPfluxapdim}$ \ergscmsq)
    \tablenotemark{i}     \dotfill    & $\hatcurPPfluxap$         \\
~~~$\langle F \rangle$ ($10^{\hatcurPPfluxavgdim}$ \ergscmsq) 
    \tablenotemark{i}     \dotfill    & $\hatcurPPfluxavg$        \\[-2ex]
\enddata
\tablenotetext{a}
    {Reference epoch of mid transit that minimizes the
    correlation with the orbital period.}
\tablenotetext{b}
    {Total transit duration, time between first to last contact.}
\tablenotetext{c}
    {Ingress/egress time, time between first 
	and second, or third and fourth contact.
	}
\tablenotetext{d}
    {Adopted from the tabulations by
    \cite{claret:2004} according to the spectroscopic (SME) parameters
    listed in \reftabl{stellar}.}
\tablenotetext{e}
    {Lagrangian orbital parameters derived from the global modeling, and
    primarily determined by the RV data.}
\tablenotetext{f}
    {The contribution of the intrinsic stellar jitter
    and possible instrumental errors
    that needs to be added in quadrature
    to the calculated RV uncertainties
    so that $\chi^{2}/{\rm dof} = 1$ in the joint fit.}
\tablenotetext{g} 
	{Correlation coefficient between 
	the planetary mass and radius.}
\tablenotetext{h}
   {The Safronov number is given by \citet{hansen:2007} as $\Theta =
   \frac{1}{2}(V_{\rm esc}/V_{\rm orb})^2 = (a/\rpl)(\mpl/\mstar)$ .}
\tablenotetext{i}
	{Stellar irradiation flux per unit surface area at periastron,
	apastron and time-averaged over the orbit, respectively.}
\end{deluxetable}

\section{Discussion}
\label{sec:discussion}

\subsection{Properties of \hatcurb{}}

\begin{figure*}[!ht]
\plotone{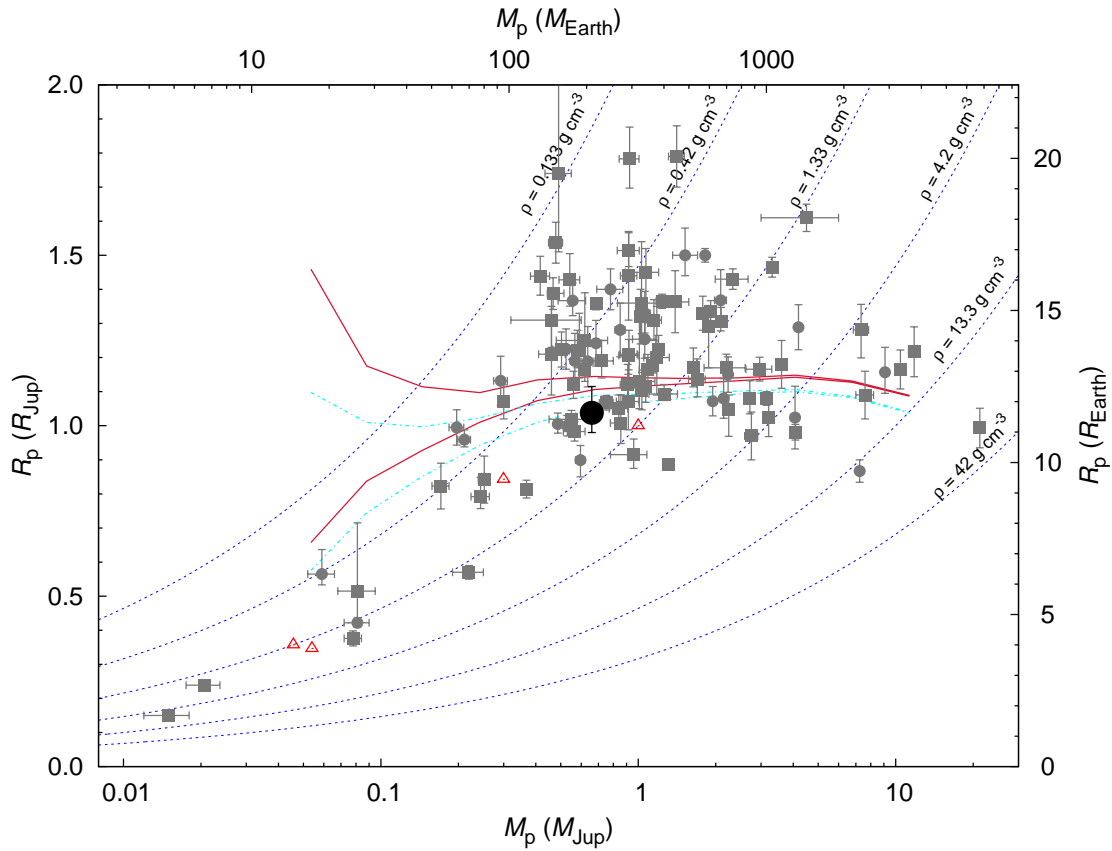}
\caption{
    Mass--radius diagram showing \hatcurb{} (solid black circle),
    other HATNet planets (solid gray circles),
    other known transiting exoplanets (solid gray squares),
    and Solar System gas giants (empty red triangles).
    Overlaid are \citet{fortney:2007} planetary isochrones interpolated
    to the solar equivalent semi-major axis of \hatcurb{} for ages of
    1 Gyr (solid crimson lines) and 4 Gyr (dashed-dotted cyan lines)
    and core masses of 0 and 10 \mearth{} (upper and lower pairs of lines
    respectively). Isodensity curves are shown for 0.133, 0.42,
    1.33 (Jupiter density), 4.2, 13.3, and 42 \gcmc (dashed lines).
\label{fig:exomr}}
\end{figure*}

\reffigl{exomr} presents the currently known transiting exoplanets and
Solar System gas planets on a mass---radius diagram, with \hatcurb{}
highlighted. Also shown are the planetary isochrones of
\citet{fortney:2007} interpolated to the insolation of \hatcur{}
at the orbit of \hatcurb{}. Taking into consideration the age
established in \refsecl{stelparam}, the planetary parameters are
consistent with a hot Jupiter with a 10 \mearth{} core in a 4 Gyr
old system.

\begin{figure*} [!ht]
\plottwo{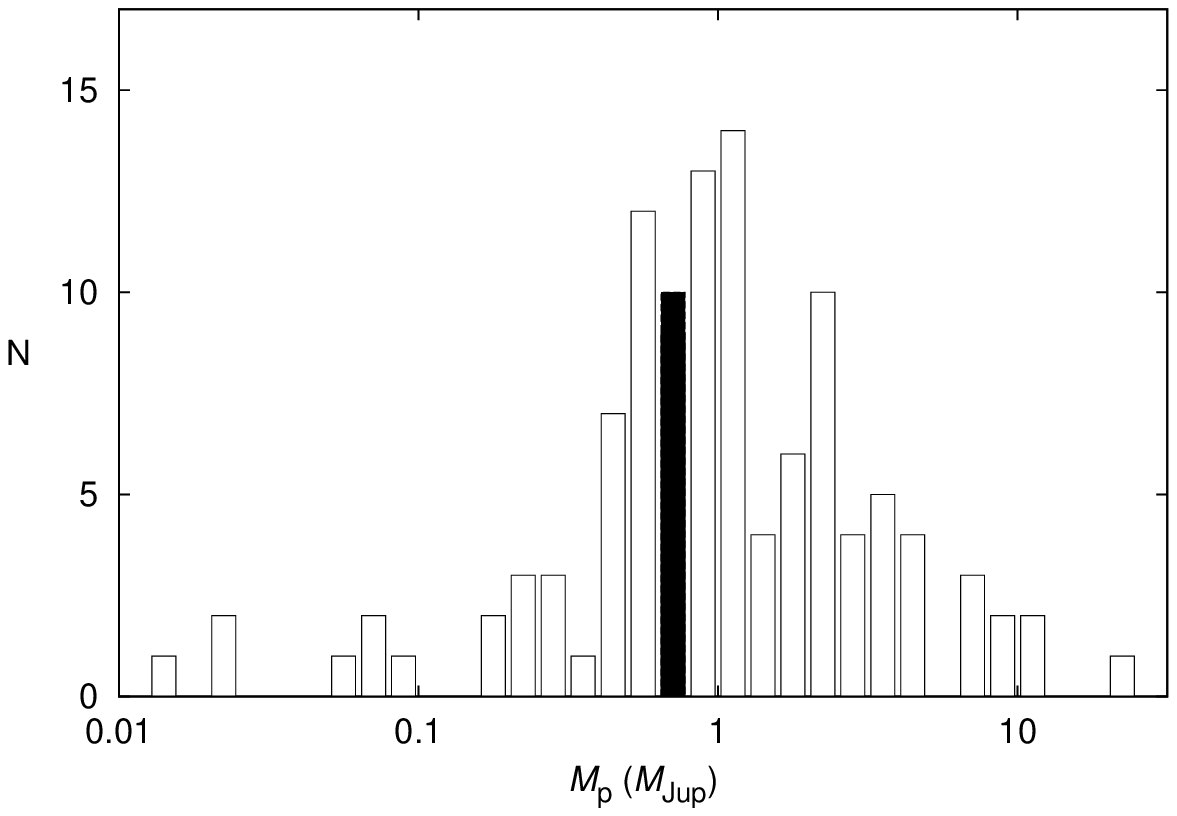}{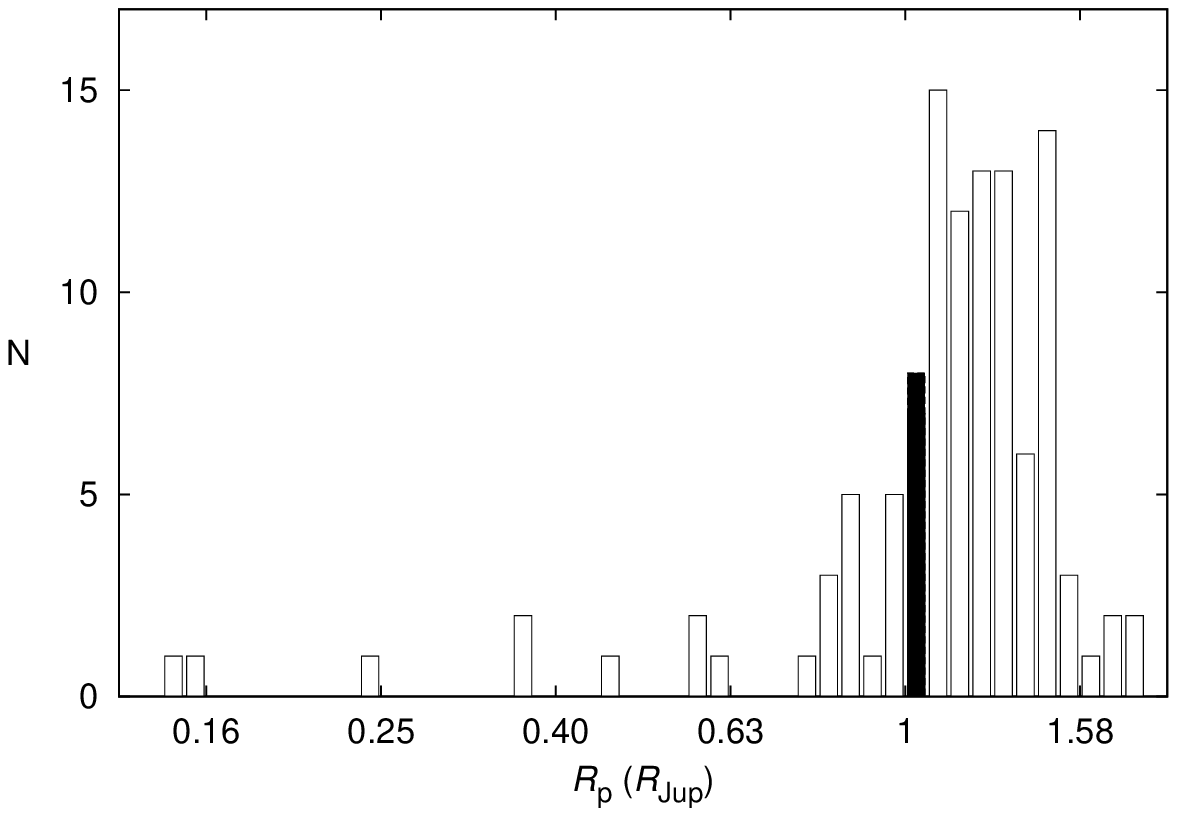}
\plottwo{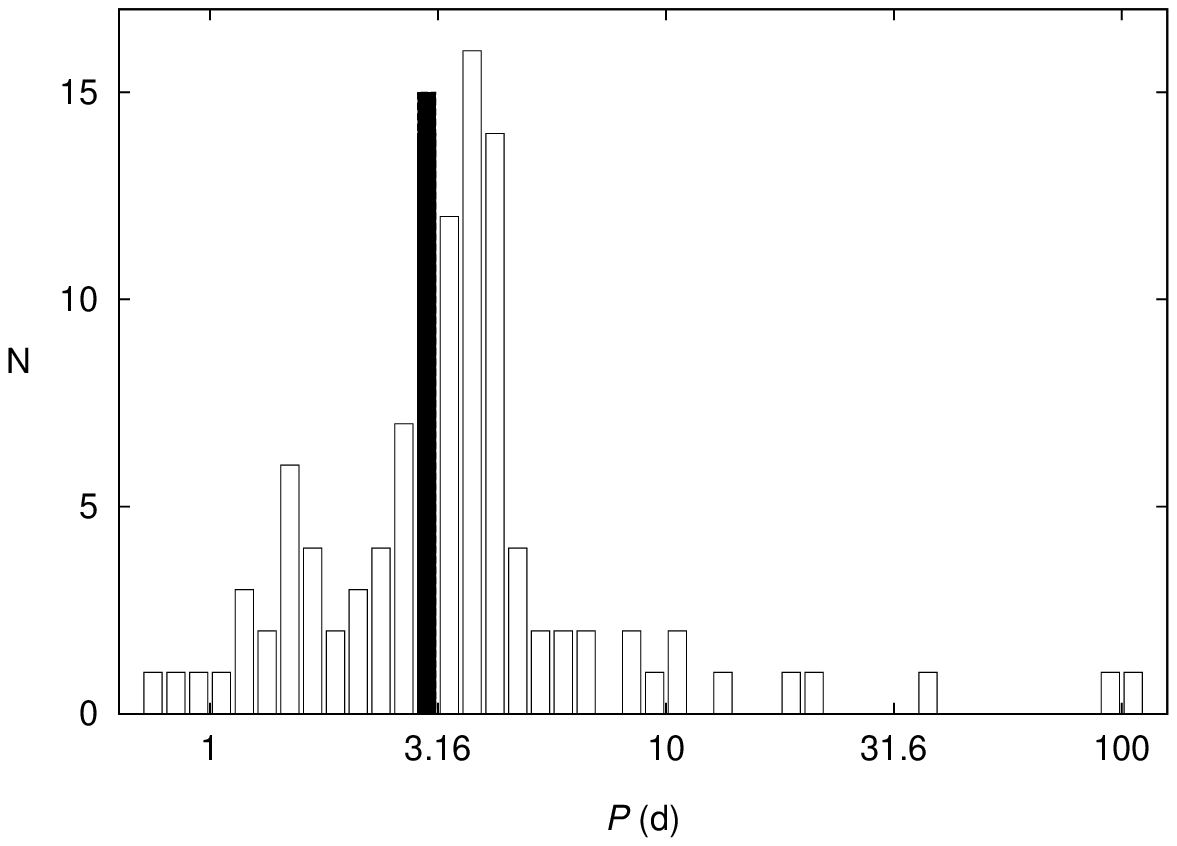}{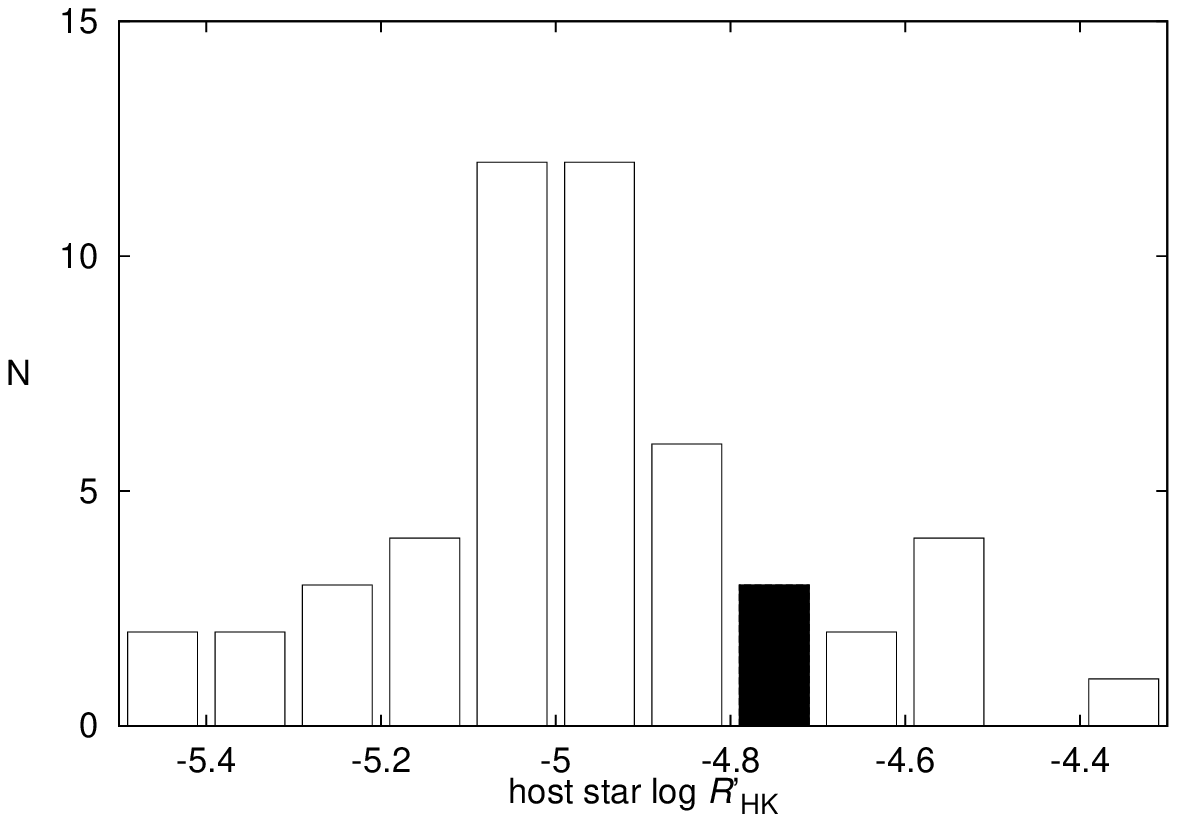}
\caption{
    Distribution of planetary and stellar parameters for 
    transiting exoplanets known to date.
    Horizontal axes and bins are logarithmic except for $\log R'_\mathrm{HK}$.
    Vertical axes show the number of planets in each bin,
    and the bin containing \hatcurb{} has solid filling.
	{\em Top left panel:} histogram of planetary mass
	in Jupiter masses, with logarithmic bin size 0.2.
	{\em Top right panel:} histogram of planetary radius
	in Jupiter radii, with logarithmic bin size 0.025.
	{\em Bottom left panel:} histogram of period in days,
	with logarithmic bin size 0.05.
	{\em Bottom right panel:} histogram of $\log R'_\mathrm{HK}$,
	with bin size 0.1. This index has only been reported for 52 host stars.
\label{fig:exohist}}
\end{figure*}

\hatcurb{} can be seen to lie inside the large
accumulation of planets with similar masses and radii.
To further compare it to other Hot Jupiters, in \reffigl{exohist} we
present histograms of mass, radius and period
for the 112
transiting exoplanets confirmed to date.

When comparing these parameters, we note that
there is only one transiting exoplanet known that is more massive
and has a smaller radius and a smaller period than \hatcurb{}:
this is HAT-P-20b with 7.246 \mjup{}, 0.867 \rjup{} on a 2.875 d orbit
\citep{bakos:2010b}.
This means
that \hatcurb{} is less inflated than other planets of similar mass
and orbital period, possibly due to a larger than average core.

Regarding eccentricity, there are 31 transiting exoplanets known under
8 \mjup{} with an orbital period within 0.5 d of that of \hatcurb{},
out of which 8 -- more than a quarter of them -- are thought to be
eccentric.
This hints that there is a possibility
for the orbit of \hatcurb{} to be eccentric as well,
justifying our choice not to fix eccentricity to zero in \refsecl{analysis}.
Future observations of radial velocity or occultation
timing would be required to determine whether the orbit is indeed
eccentric.

The impact parameter of \hatcurb{} is unusually large. 
As \citet{ribas:2008} pointed out, such a near grazing transit
has the advantage of its depth and duration being more sensitive
to the presense of further planetary companions on inclined orbits.
This makes \hatcur{} a promising target for transit timing variation
and transit duration variation studies.

\citet{knutson:2010} found a strong negative correlation between
chromospheric activity of the host star and temperature inversion in
the planetary atmosphere.  However, since early type stars dominate
magnitude limited surveys, cool, that is, active planetary hosts are
rare.  The bottom right panel of \reffigl{exohist} shows that \hatcur{}
is relatively active compared to known planetary hosts for which $\log
R'_\mathrm{HK}$ has been reported, making it an exciting target for
{\em Spitzer Space Telescope} to test this correlation.

\subsection{Correlation of planetary parameters with host star metallicity}

The relation between host star metallicity (denoted as $\fehstar$ for
clarity, not to be confused with the assumed metal content of the planet) 
and planetary composition was studied by \citet{guillot:2006}.  
A positive correlation, with
Pearson correlation coefficient $r=0.78$, was found between the
inferred mass of the planetary core and stellar metallicity for the
seven transiting exoplanets known at that time with 
positive
inferred
core mass.  The idea is that planets have formed from the same cloud as
their host stars, 
and
therefore their metal content should correlate. 
However, it is not clear how stellar metallicity is connected to
planetary metallicity, especially because a larger rocky core is likely
to accrete more gas during the planet's formation.

\citet {burrows:2007} also investigated this relation, based on
12 transiting exoplanets known at the time. They
used an atmospheric opacity dependent core mass model to
explain radius anomalies. 
They also found a strong
correlation between host star metallicity and inferred core mass, but the
correlation coefficient was not reported.

\citet{enoch:2010} found that there is a
strong negative correlation with $r=-0.53$ between \fehstar\ and \rpl\ for
the 18 known transiting exoplanets below 0.6 \mjup{},
whereas this correlation is negligible for more massive planets.
This can be explained by noticing that
the theoretical planet models of \citet{fortney:2007}, \citet{bodenheimer:2003},
and \citet{baraffe:2008} all suggest that the radius of a planet is more sensitive
to its composition for low mass planets than it is for more massive ones.

\begin{deluxetable*}{llllll}
\tablewidth{0pc}
\tabletypesize{\scriptsize}
\tablecaption{
    Parameters of 30 transiting exoplanets between 0.3 \mjup{} and 0.8 \mjup{} in increasing order of mass
    \label{tab:planets}
}
\tablehead{
  \multicolumn{1}{c}{name}                 &
  \multicolumn{1}{c}{$ \mpl (\mjup) $}     &
  \multicolumn{1}{c}{$ \rpl (\rjup) $}     &
  \multicolumn{1}{c}{$ \teff (\mathrm K)$} &
  \multicolumn{1}{c}{$ \fehstar $}         &
  \multicolumn{1}{c}{reference}
}
\startdata
WASP-21b          & $0.3\pm0.01$              & $1.07\pm0.05$             & $1262\pm31$          & $-0.4\pm0.1$    & \citet{277} \\
HD 149026b        & $0.368^{+0.013}_{-0.014}$ & $0.813^{+0.027}_{-0.025}$ & $1792^{+44}_{-32}$   & $+0.36\pm0.05$  & \citet{55},\\&&&&&\citet{73} \\
Kepler-7b         & $0.416^{+0.036}_{-0.035}$ & $1.439^{+0.058}_{-0.056}$ & $1565^{+31}_{-30}$   & $+0.11\pm0.03$  & \citet{161},\\&&&&&\citet{284} \\
WASP-13b          & $0.46^{+0.056}_{-0.05}$   & $1.21^{+0.14}_{-0.12}$    & $1417^{+62}_{-58}$   & $+0.0\pm0.2$    & \citet{244} \\
Kepler-8b         & $0.46\pm0.14$             & $1.31^{+0.076}_{-0.08}$   & $1628^{+52}_{-53}$   & $-0.055\pm0.03$ & \citet{162},\\&&&&&\citet{284} \\
CoRoT-5b          & $0.467^{+0.047}_{-0.024}$ & $1.388^{+0.046}_{-0.047}$ & $1438\pm39$          & $-0.25\pm0.06$  & \citet{35} \\
WASP-31b          & $0.478\pm0.03$            & $1.54\pm0.06$             & $1568\pm33$          & $-0.19\pm0.09$  & \citet{317} \\
WASP-11/HAT-P-10b & $0.487\pm0.018$           & $1.005^{+0.032}_{-0.027}$ & $1020\pm17$          & $+0.13\pm0.08$  & \citet{20} \\
WASP-17b          & $0.49^{+0.059}_{-0.056}$  & $1.74^{+0.26}_{-0.23}$    & $1662^{+113}_{-110}$ & $-0.25\pm0.09$  & \citet{249} \\
WASP-6b           & $0.503^{+0.019}_{-0.038}$ & $1.224^{+0.051}_{-0.052}$ & $1194^{+58}_{-57}$   & $-0.20\pm0.09$  & \citet{239} \\
HAT-P-1b          & $0.524\pm0.031$           & $1.225\pm0.059$           & $1306\pm30$          & $+0.21\pm0.03$  & \citet{6},\\&&&&&\citet{55} \\
HAT-P-17b         & $0.53\pm0.019$            & $1 293\pm0.03$            & $787\pm15$           & $+0.0\pm0.08$   & \citet{293} \\
WASP-15b          & $0.542\pm0.05$            & $1.428^{+0.077}_{-0.077}$ & $1652\pm28$          & $-0.17\pm0.11$  & \citet{247} \\
OGLE-TR-111b      & $0.55\pm0.1$              & $1.019^{+0.026}_{-0.026}$ & $1025^{+26}_{-25}$   & $+0.19\pm0.07$  & \citet{168},\\&&&&&\citet{6},\\&&&&&\citet{211} \\
HAT-P-4b          & $0.556\pm0.068$           & $1.367^{+0.052}_{-0.044}$ & $1686^{+30}_{-26}$   & $+0.24\pm0.08$  & \citet{11},\\&&&&&\citet{6},\\&&&&&\citet{312} \\
WASP-22b          & $0.56\pm0.02$             & $1.12\pm0.04$             & $1430\pm30$          & $-0.05\pm0.08$  & \citet{273}  \\
XO-2b             & $0.566^{+0.055}_{-0.055}$ & $0.983^{+0.029}_{-0.028}$ & $1319^{+24}_{-23}$   & $ 0.44\pm0.04$  & \citet{6},\\&&&&&\citet{55} \\
HAT-P-25b         & $0.567\pm0.022$           & $1.19^{+0.081}_{-0.056}$  & $1202\pm36$          & $+0.31\pm0.08$  & \citet{quinn:2010} \\
WASP-25b          & $0.58\pm0.04$             & $1.22^{+0.06}_{-0.05}$    & $1212\pm35$          & $-0.07\pm0.1$   & \citet{enoch:2010} \\
WASP-34b          & $0.59\pm0.01$             & $1.22^{+0.11}_{-0.08}$    & $1250\pm30$          & $-0.02\pm0.1$   & \citet{319} \\
HAT-P-3b          & $0.596^{+0.024}_{-0.026}$ & $0.899^{+0.043}_{-0.049}$ & $1127^{+49}_{-39}$   & $+0.27\pm0.04$  & \citet{8},\\&&&&&\citet{6} \\
HAT-P-28b         & $0.636\pm0.037$           & $1.189^{+0.102}_{-0.075}$ & $1371\pm50$          & $+0.12\pm0.08$  & \citet{316} \\
HAT-P-27b         & $0.660\pm0.033$           & $1.038^{+0.077}_{-0.058}$ & $1207\pm41$          & $+0.29\pm0.1$   & this paper \\
HAT-P-24b         & $0.685\pm0.033$           & $1.242\pm0.067$           & $1637\pm42$          & $-0.16\pm0.08$  & \citet{296} \\
HD 209458b        & $0.685^{+0.015}_{-0.014}$ & $1.359^{+0.016}_{-0.019}$ & $1449\pm12$          & $+0.01\pm0.03$  & \citet{137},\\&&&&&\citet{6} \\
Kepler-6b         & $0.62^{+0.025}_{-0.028}$  & $1.164^{+0.025}_{-0.017}$ & $1459^{+25}_{-24}$   & $+0.34\pm0.04$  & \citet{160},\\&&&&&\citet{284} \\
OGLE-TR-10b       & $0.62\pm0.14$             & $1.25^{+0.14}_{-0.12}$    & $1481^{+71}_{-55}$   & $+0.15\pm0.15$  & \citet{6} \\
CoRoT-4b          & $0.72\pm0.08$             & $1.19^{+0.06}_{-0.05}$    & $1074\pm19$          & $+0.05\pm0.07$  & \citet{34} \\
TrES-1b           & $0.752^{+0.047}_{-0.046}$ & $1.067^{+0.022}_{-0.021}$ & $1140^{+13}_{-12}$   & $+0.02\pm0.05$  & \citet{6} \\
HAT-P-9b          & $0.780\pm0.090$           & $1.40\pm0.06$             & $1530\pm40$          & $+0.12\pm0.2$   & \citet{18} \\[-2ex]
\enddata
\end{deluxetable*}

In this subsection, we examine further the
correlation between host star metallicity and planetary mass or planetary
radius. We use a substantially expanded sample of 
30 known transiting exoplanets with masses
between 0.3 \mjup{} and 0.8 \mjup{}, see \reftabl{planets}.
The upper limit is selected to exclude more massive planets whose radius
is expected to depend less on the composition, see above.
We explain the role of the lower limit and the effect of the two lowest mass
planets in \reftabl{planets} later.

The null hypothesis is that the host star metallicity
and the selected planetar parameter are independent.  The alternative
hypothesis is that they are related by some underlying phenomenon.  A
false positive, also known as an error of the first kind, 
is rejection
of the null hypothesis in spite of it being true.  We implement three
independent statistical methods to estimate the false positive
probability: $t$-test, bootstrap technique and $F$-test.  We denote the
probability estimates by $p_1$, $p_2$ and $p_3$, respectively.  This is
the statistical significance of the correlation: the lower this
probability is, the more confidently the null hypothesis (i.e., no
correlation) can be rejected.

For the $t$-test, we assume that the investigated parameters have
normal distribution.  For each set of data pairs,
we calculate the $t$
value from the correlation coefficient $r$ and sample size $n$ using
the formula
$$ t = r \sqrt {\frac {n-2}{1-r^2}}\;.$$
The conditional distribution of this variable given the null hypothesis
is Student's $t$ distribution with $n-2$ degrees of freedom
\citep[p.~640]{press:1992}. Then the estimate $p_1$ for
false positive probability is determined
by looking up the two-tailed probability of this distribution
yielding larger absolute value that the one measured.
For comparison, we also performed the $t$-test for the samples
and parameters studied by \citet{guillot:2006} and \citet{enoch:2010}.
The resulting values are 
listed in \reftabl{corr}.

For the sample set of 
\reftabl{planets},
we also implement the bootstrap resampling technique
\citep{efron:2003}.  This has the advantage that no assumption about
the {\em a priori} distribution of the data is necessary.  
To perform
bootstrap resampling, consider the data $(x_1, y_1), (x_2, y_2),
\ldots, (x_n, y_n)$, where $x_i$ is the host star metallicity, and
$y_i$ is the corresponding planetary mass or radius.  Again, we would
like to calculate an estimate $p_2$ of the probability that a sample of
similar distribution, but independent parameters for each pair, has a
correlation coefficient that exceeds that of our measurements in
absolute value.  For this, we build 10\,000\,000 sample sets of $n$
pairs by drawing $x$ and $y$ values independently with replacements
from the set of measured $x$ and $y$ values, respectively.  The
percentile rank of the absolute value of the correlation coefficient
for the measured data among the random samples gives our estimates
$p_2$, listed in \reftabl{corr}.

Finally, we test these correlations with an additional method, the $F$-test
\citep[see e.g.][p.~100]{lupton:1993}.  This requires that the null
hypothesis (no correlation) be nested in the tested hypothesis (linear
correlation), which indeed is the case.  Let RSS$_1$ denote the
residual sum of squares for the best fit of the null hypothesis, that
is, the variance of $y_i$ about its mean, and RSS$_2$ denote the
residual sum of squares of the linear fit.  The no correlation model
has $l_1=1$ free parameters: the mean, whereas the linear fit has
$l_2=2$.  The conditional distribution of
$$ F = \frac {\frac {\mathrm{RSS}_1 - \mathrm{RSS}_2} {l_2-l_1}} 
{\frac {\mathrm{RSS}_2}{n-l_2}} $$
given the null hypothesis is the $F$-distribution with $(l_2-l_1,n-l_2)$
degrees of freedom.  This enables us to calculate $p_3$, the third
estimate for the false positive probability.

The estimates $p_1$, $p_2$, $p_3$ given by the three methods
are listed in \reftabl{corr}. 
For each correlation, they coincide up to the uncertainty of the
methods. 
The values reflect the significant
$\fehstar\textrm{---}M_{core}$ and $\fehstar\textrm{---}\rpl$
correlations found by \citet{guillot:2006} and \citet{enoch:2010}.

As for the 30 planets listed in \reftabl{planets},
it is important to note that the correlations depend strongly on the
choice of the lower mass limit. The two least massive planets in the table
are WASP-21b with a mass of $0.3\;\mjup$, very low host star metallicity of $-0.4$,
and average radius of $1.07\;\rjup$;
and the dense HD 149026b with a mass of $0.368\;\mjup$, 
high host star metallicity of $+0.36$,
and low radius of $0.813\;\rjup$.
Increasing the lower mass limit for our sample first excludes WASP-21b,
which would much support the positive $\fehstar$---$\mpl$ correlation
with its low mass and low host star metallicity.
Further increasing the limit then excludes HD 149026b, which would
much weaken it with its low mass and high host star metallicity.
To have an unbiased result, outliers cannot be excluded
without a justified reason, therefore we need to compare the false positive
probabilities of the three nested samples. They scatter between 15\% and 58\%,
neither supporting, nor rejecting a $\fehstar$---$\mpl$ correlation.

Similarily, both WASP-21b and HD 149026b have a strong effect
on the negative $\fehstar$---$\rpl$ correlation, because of the extreme
value of their host star metallicities. In this case, we see that the
maximum of the false positive probabilities is $0.44\%$, therefore
this correlation is statistically significant for all our choices
of lower mass limits. This is at least a fivefold improvement over the sample
investigated by \citet{enoch:2010}, due to the larger sample size.

\begin{deluxetable*}{r@{}c@{}lccr@.lr@.lr@.lr@.l}
\tablewidth{0pc}
\tabletypesize{\scriptsize}
\tablecaption{
    Correlation between the host star metallicity and planetary
    parameters for known transiting exoplanets
    \label{tab:corr}
}
\tablehead{
    \multicolumn{3}{c}{Restriction}          &
    \multicolumn{1}{c}{$n$\tablenotemark{a}} &
    \multicolumn{1}{c}{Planetary}            &
    \multicolumn{2}{c}{$r$\tablenotemark{b}} &
    \multicolumn{2}{c}{$p_1$\tablenotemark{c}}&
    \multicolumn{2}{c}{$p_2$\tablenotemark{c}}&
    \multicolumn{2}{c}{$p_3$\tablenotemark{c}}\\
    \multicolumn{3}{c}{on planets}
&&   \multicolumn{1}{c}{parameter}
}
\startdata
       $0<$  & $M_\mathrm{core}$ &  &  7 & $M_\mathrm{core}$ &  0&78\tablenotemark{d} &  3&9\%      \\
             &  \mpl & $<0.6$ \mjup & 18 & \rpl              & -0&53\tablenotemark{e} &  2&4\%       \\
0.3  $\mjup\leqslant$ & \mpl & $\leqslant0.8$ \mjup & 30 & \mpl &  0&270  & 15&0\% & 15&0\%  & 15&0\% \\
0.35 $\mjup\leqslant$ & \mpl & $\leqslant0.8$ \mjup & 29 & \mpl &  0&106  & 58&4\% & 58&2\%  & 58&4\%  \\
0.4  $\mjup\leqslant$ & \mpl & $\leqslant0.8$ \mjup & 28 & \mpl &  0&247  & 20&4\% & 20&4\%  & 20&4\%   \\
0.3  $\mjup\leqslant$ & \mpl & $\leqslant0.8$ \mjup & 30 & \rpl & -0&505  &  0&44\% & 0&43\% &  0&44\%   \\
0.35 $\mjup\leqslant$ & \mpl & $\leqslant0.8$ \mjup & 29 & \rpl & -0&620  &  0&03\% & 0&03\% &  0&034\%   \\
0.4  $\mjup\leqslant$ & \mpl & $\leqslant0.8$ \mjup & 28 & \rpl & -0&575  &  0&14\% & 0&14\% &  0&14\%     \\[-2ex]
\enddata
\tablenotetext{a}{sample size}
\tablenotetext{b}{correlation coefficient}
\tablenotetext{c}{estimates for false positive probability given by $t$-test, bootstrap method and $F$-test, respectively}
\tablenotetext{d}{reported by \citet{guillot:2006}}
\tablenotetext{e}{reported by \citet{enoch:2010}}
\end{deluxetable*}

\subsection{Dependence on planetary equilibrium temperature}

Other factors like insolation are likely to influence planetary radius
as well, see e.g.~\cite{fortney:2007}, \cite{kovacs:2010}, \cite{enoch:2010},
and \cite{faedi:2011}. To
further investigate this relation, we compare two models: for null hypothesis,
we accept the linear planetary radius---host star metallicity relation
of the previous section:
\begin {equation}
\tilde \rpl^\mathrm{I} = R_0^\mathrm{I} + \alpha^\mathrm{I} \cdot\fehstar.
\end {equation}

The second model -- alternative hypothesis -- is similar to that
of \citet {enoch:2010}, accounting for the equilibrium temperature
$T_\mathrm{eq}$ in addition to the host star metallicity:
\begin {equation}
\label{eq:secondmodel}
\tilde \rpl^\mathrm{II} = R_0^\mathrm{II} + \alpha^\mathrm{II} \cdot\fehstar + \beta^\mathrm{II} \cdot T_\mathrm{eq}.
\end {equation}
The equilibrium temperature of the planet is calculated from
the time-averaged stellar flux on its orbit,
assuming gray body spectrum for the planets,
and neglecting tidal and other heating mechanisms.
For simplicity, we now include all 30 planets of \reftabl {planets} in our models.
With the best fit parameters, the two models are
\begin{eqnarray*}
\tilde \rpl^\mathrm{I}  & = & 1.235\,\rjup{}-0.478\,\rjup{} \cdot \fehstar,\\
\tilde \rpl^\mathrm{II} & = & 0.690\,\rjup{}-0.431\,\rjup{} \cdot \fehstar + 0.398 \,\rjup \cdot \frac {T_\mathrm{eq}}{1000\,\mathrm K}.
\end{eqnarray*}

To quantify the statistical significance,
we apply the $F$-test again.
The resulting false positive probability is 0.18\%.
This means that once we accept the dependence of planetary radius
on host star metallicity, then the probability of such a correlation
with planetary equilibrium temperature if they were not physically
related is only 0.18\%. This strongly supports the three-parameter
linear fit model.
For reference, the residual sums of
squares for the one, two and three-parameter fits of the
\fehstar---\rpl\ data are $1.13\;\rjup^2$, $0.84\;\rjup^2$ and $0.58\;\rjup^2$, respectively.

\begin{figure*} [!ht]
\plotone{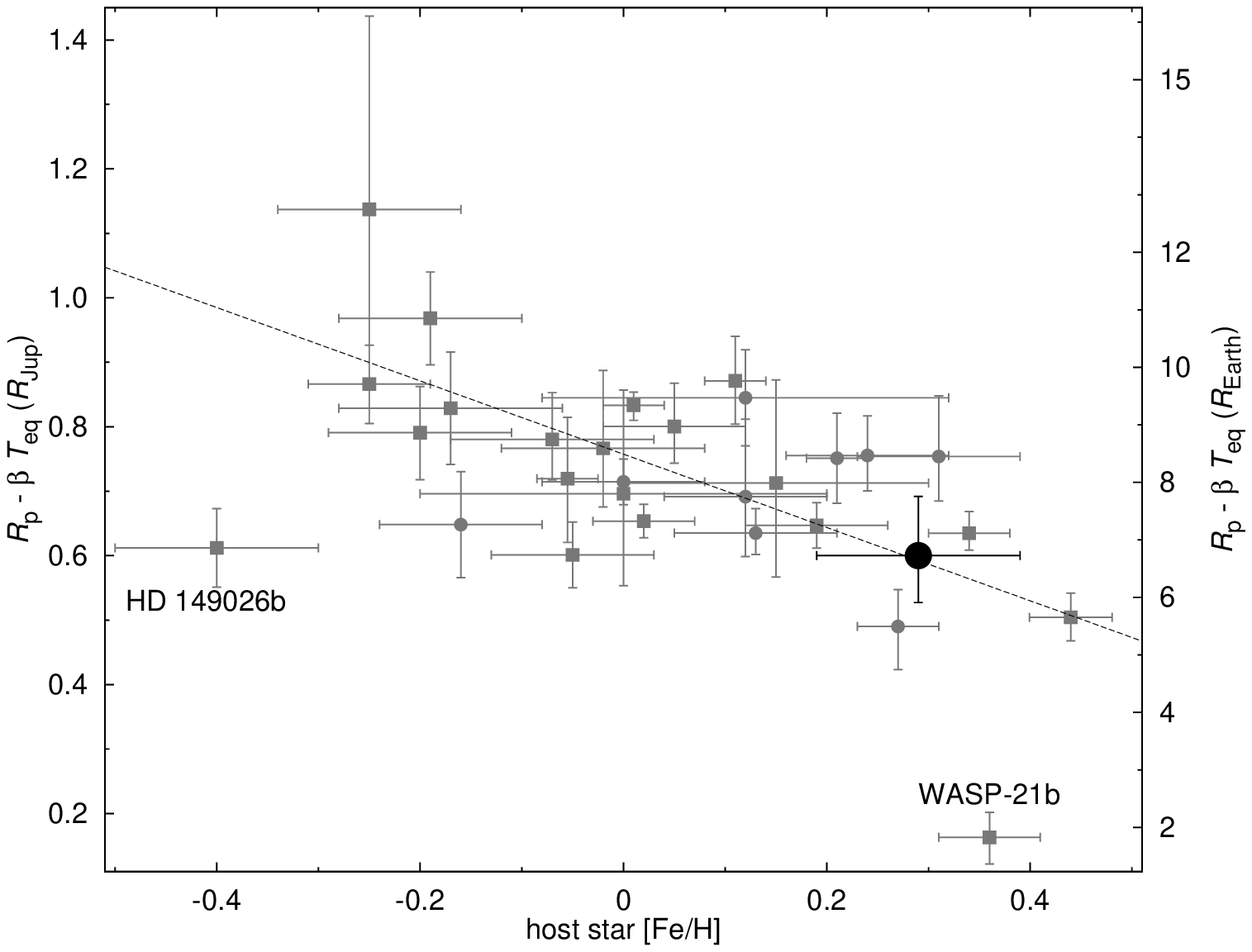}
\caption{
	Planetary radius corrected for linear equilibrium temperature dependence
	in Jupiter radii versus host star metallicity for
	the 30 known transiting exoplanets with masses between 0.3 \mjup{} and
	0.8 \mjup{}, including
    \hatcurb{} (solid black circle),
    other HATNet planets (solid gray circles),
    and other known transiting exoplanets (solid gray squares,
    WASP-21b and HD 149026b labeled);
    with best linear fit overlaid (dashed line).
\label{fig:exocorr}}
\end{figure*}

\reffigl{exocorr} presents $\rpl-\beta^\mathrm{II}\cdot T_\mathrm{eq}$
versus metallicity for the 30 planets.  Equation (\ref{eq:secondmodel})
predicts this quantity to be
$R_0^\mathrm{II}+\alpha^\mathrm{II}\cdot\fehstar$, which is also plotted. 
\hatcurb{} apparently follows the model's prediction.  For reference,
the correlation coefficient between the displayed transformed variables
is now $r^\mathrm{II}=-0.536$, which has a larger absolute value than
$r^\mathrm{I}=-0.505$ between $\rpl$ and metallicity, as expected.

This analysis supports the statement that planetary radius depends on
equilibrium temperature in addition to host star metallicity, as found
by \citet {enoch:2010} and \citet{faedi:2011}.  However, this
correlation does not imply that insolation itself would inflate
planets: the underlying phenomenon could be related to anything
correlated to equilibrium temperature, or equivalently, orbital radius. 
For instance, \citet {batygin:2010} suggest that it is Ohmic
dissipation in the interior of the planet that inflates hot Jupiters. 
This theory is further supported by \citet {laughlin:2011}.

Altogether, \hatcurb{} is an important addition to the growing sample
of low-mass Jupiters. It orbits a metal rich star, and supports the
suggested correlations between host star metallicity,
planetary equilibrium temperature, and planetary radius. Also,
\hatcur{} is chromospherically active, providing an excellent case
for refining the confidence level of the hypothesized correlation
between stellar activity and planetary temperature inversion.

\acknowledgements 

HATNet operations have been funded by NASA grants NNG04GN74G,
NNX08AF23G and SAO IR\&D grants.  Work of G.\'A.B.~was
supported by the Postdoctoral Fellowship of the NSF Astronomy and
Astrophysics Program AST-0702843.
G.T.~acknowledges partial support from NASA grant NNX09AF59G.
A.J.~acknowledges support from Fondecyt project 1095213, BASAL CATA
PFB-06, FONDAP CFA 15010003, MIDEPLAN ICM Nucleus P07-021-F and Anillo
ACT-086.
G.K.~thanks the Hungarian Scientific Research Foundation (OTKA)
for support through grant K-81373.
L.L.K.~is supported by the ``Lend\"ulet'' Young Researchers Program
of the Hungarian Academy of Sciences
and the Hungarian OTKA grants K76816, K83790 and MB08C 81013.
Tam\'as Szalai (Univ.~of Szeged) is acknowledged for his
assistance during the ANU 2.3 m observations.
We acknowledge partial support also from the Kepler Mission under NASA
Cooperative Agreement NCC2-1390 (D.W.L., PI).
This research has made use of Keck telescope time
granted through NOAO (A201Hr) and NASA (N018Hr, N167Hr).
We also thank Mount Stromlo Observatory and Siding Spring Observatory
for the ANU 2.3 m telescope time.

\end{document}